\documentclass[11pt]{article}
\usepackage{amssymb}
\usepackage{amsfonts}
\usepackage{graphicx}
\usepackage{amsmath}
\usepackage{hyperref}

\makeatletter \@addtoreset{equation}{section} \makeatother

\newtheorem{theorem}{Theorem}
\newtheorem{lemma}{Lemma}

\newtheorem{remark}{Remark}

\newtheorem{proposition}{Proposition}

\begin{document}

\title{Fluctuations of linear eigenvalue statistics
of $\beta$ matrix models in the multi-cut
regime}
\author{ M. Shcherbina
\\
Institute for Low Temperature Physics Ukr.Ac.Sci, Kharkov, Ukraine.\\
E-mail:shcherbi@ilt.kharkov.ua
}

\date{}

\maketitle

\begin{abstract}
We study the asymptotic expansion in $n$ for the partition function of  $\beta$ matrix models with real analytic
potentials in the multi-cut regime up to the $O(n^{-1})$ terms. As a result, we find the limit of the generating
functional of linear eigenvalue statistics  and the expressions for the expectation and the variance of linear
 eigenvalue statistics, which in the general case contain the quasi periodic in $n$ terms.
\end{abstract}

\section{Introduction and main results}\label{s:1}

In this paper we consider a class of  distributions in $\mathbb{R}^n$ of the form
\begin{equation}\label{p(la)}
p_{n,\beta}(\lambda_1,...,\lambda_{n})=Q_{n,\beta}^{-1}[V]\prod_{i=1}^n
e^{-n\beta V(\lambda_i)/2}\prod_{1\le i<j\le
n}|\lambda_i-\lambda_j|^\beta=Q_{n,\beta}^{-1}e^{\beta H(\lambda_1,\dots,\lambda_n)/2},
\end{equation}
where the function $H$, which we call Hamiltonian to stress the analogy with statistical mechanics,
and the normalizing constant $Q_{n,\beta}[V]$ (called the partition function) have the form
\begin{align}\label{H}
&H(\lambda_1,\dots,\lambda_n)=-n\sum_{i=1}^n
V(\lambda_i)+\sum_{i\not=j}\log|\lambda_i-\lambda_j|,\\
&Q_{n,\beta}[V]=\int e^{\beta H(\lambda_1,\dots,\lambda_n)/2}d\lambda_1\dots d\lambda_{n}.
\label{Q_n}\end{align}
We denote also
\begin{align}\label{E}
    \mathbf{E}_{n,\beta}\{(\dots)\}=&\int(\dots)p_{n,\beta}(\lambda_1,...,\lambda_{n})d\lambda_1,\dots
    d\lambda_{n},\\
\label{p_nl}
p^{(l)}_{n,\beta}(\lambda_1,...,\lambda_l)=&
\int_{\mathbb{R}^{n-l}} p_{n,\beta}(\lambda_1,...\lambda_l,\lambda_{l+1},...,\lambda_{n})
d\lambda_{l+1}...d\lambda_{n}
\end{align}
the corresponding expectation and the $l$th marginal densities (correlation functions) of (\ref{p(la)}).
The function $V$ in (\ref{H}), called the potential, is a real valued H\"{o}lder function satisfying the condition
\begin{equation}\label{condV}
V(\lambda )\ge 2(1+\epsilon )\log(1+ |\lambda |).
\end{equation}
  Such distributions can be considered
for any $\beta>0$, but the cases $\beta=1,2,4$ are especially important, since they
correspond  to real symmetric, hermitian, and symplectic
matrix models respectively.

Since the papers \cite{BPS:95,Jo:98} it is known that
if $V$ is a H\"{o}lder function, then
\[n^{-2}\log Q_{n,\beta}[V]=\frac{\beta}{2}\mathcal{E}[V]+O(\log n/n),\]
where
\begin{equation}\label{E_V}
\mathcal{E}[V]=\max_{m\in\mathcal{M}_1}\bigg\{
L[dm,dm]-\int V(\lambda)m(d\lambda)\bigg\}=\mathcal{E}_V(m^*),
\end{equation}
and the maximizing measure $m^*$ (called the equilibrium measure) has a compact
support $\sigma:=\mathrm{supp\,}m^*$. Here and below
 we denote
\begin{align}\label{L[,]}
&L[\,dm,dm]=\int\log|\lambda-\mu|dm(\lambda) dm(\mu),\\
&L[f](\lambda)=\int\log|\lambda-\mu|f(\mu)d\mu, \quad L[f,g]=(L[f],g),
\notag\end{align}
where $(.,.)$ is a standard inner product in $L_2[\mathbb{R}]$.

If $V'$ is a H\"{o}lder function, then the equilibrium measure $m^*$ has a density
$\rho$ (equilibrium density).  The support $\sigma$ and the density $\rho$ are uniquely defined by the conditions:
\begin{equation}\label{cond_rho}\begin{array}{l}\displaystyle
v(\lambda ):=2\int \log |\mu -\lambda |\rho (\mu )d\mu -V(\lambda )=\sup v(\lambda):=v^*,\quad\lambda\in\sigma,\\
v(\lambda )\le \sup v(\lambda),\quad \lambda\not\in\sigma,\hskip
2cm\sigma=\hbox{supp}\{\rho\}.
\end{array}\end{equation}
Without loss of generality we will assume below that $\sigma\subset(-1,1)$ and $v^*=0$.

In this paper we  discuss the asymptotic expansion in $n^{-k}$ of the partition function
$ Q_{n,\beta}[V]$ and of the Stieltjes transforms of the marginal densities. The problems
of this kind appear in many fields of mathematics, e.g., statistical mechanics
 of $\log$-gases, combinatorics (graphical  enumeration), theory of orthogonal polynomials
etc (see  \cite{Er-McL:03} for the detailed and interesting discussion on the motivation
of the problem). Here we are going to discuss with more details
 the applications of the problem to the analysis of the eigenvalue distributions of random matrices.

 One of the most important problems of the eigenvalue distribution is the behavior of the random
 variables, called the linear
 eigenvalue statistics,  corresponding to the smooth test function $h$
\begin{equation}\label{lst}
  \mathcal{N}_n[h] =\sum_{i=1}^n h(\lambda_i).
\end{equation}
The result of \cite{BPS:95} gives us the main term of the expectation of
$\mathbf{E}_{n,\beta}\{\mathcal{N}_n[h]\}$ which is $n(h,\rho)$. It was also proven in \cite{BPS:95} that
the variance of $\mathcal{N}_n[h]$ tends to zero, as $n\to\infty$.
But the behavior of the fluctuations of $\mathcal{N}_n[h]$ was studied only in the case
of one-cut potentials (see \cite{Jo:98}). Even the bound for
$\mathbf{Var}_{n,\beta}\{\mathcal{N}_n[h]\}$ in the multi-cut regime till the recent time was known only for
$\beta=2$.  Thus the behavior of the characteristic functional,
corresponding to the linear eigenvalue statistics (\ref{lst}) of the test function
$h$
\begin{equation}\label{Z[h]}
    Z_{n,\beta}[h]=\mathbf{E}_{n,\beta}\Big\{e^{\beta(\mathcal{N}_n[h]-
    \mathbf{E}_{n,\beta}\{\mathcal{N}_n[h]\})/2}\Big\}=
 \frac{Q_{n,\beta}\big[V-\frac{1}{n}(h-
 \mathbf{E}_{n,\beta}\{n^{-1}\mathcal{N}_n[h]\})\big]}{Q_{n,\beta}\big[V\big]}.
\end{equation}
is one of the questions of primary interest in the random matrix theory. Since
$Z_{n,\beta}[h]$ is a ratio of two partition functions,
to study the behavior of $Z_{n,\beta}[h]$, it suffices to find the coefficients of the expansion
of $\log Q_{n,\beta}\big[V]$ up to the order $O(n^{-1})$.

Let us mention  the most important results on the expansion of $\log Q_{n,\beta}[V]$
and the correlation functions. The CLT for linear eigenvalue statistics in the one-cut regime for any $\beta$
and polynomial $V$ was proven in \cite{Jo:98}.
The expansion for the first and the second correlators for $\beta=2$ and one-cut real
analytic $V$  was constructed in \cite{APS:01}. The expansion of
$\log Q_{n,\beta}[V]$ for a one-cut polynomial $V$ and $\beta=2$ was obtained in
\cite{Er-McL:03}. The formal expansions for any
$\beta$ and polynomial $V$ were obtained in the physical papers \cite{CEy:06} and \cite{Ey:09}.
The CLT for $\beta=2$, real analytic multi-cut $V$, and  special choice $h=V$  was obtained in \cite{P:06}.
The expansion of $\log Q_{n,\beta}[V]$ up to $O(1)$ for one-cut real analytic $V$ and multi-cut
real analytic $V$ was performed in \cite{KS:10} and \cite{S:11} respectively. The complete asymptotic expansion
of the partition function and all the correlators for  one-cut real analytic $V$ and any $\beta$
was constructed in \cite{BG:11}. It worth to mention that the papers \cite{Jo:98,KS:10,BG:11} are based on the same method,
the first version of which was proposed in \cite{Jo:98}. The method is based on the analysis of the  first loop equation
by the methods of the perturbation theory, where the results of \cite{BPS:95} give zero order approximation.
The subsequent papers \cite{KS:10,BG:11} simplified and developed the method of \cite{Jo:98}. This allowed to the authors
to extend the method to non-polynomial $V$ (see \cite{KS:10}), and to apply it to the loop equations of higher orders
(see \cite{BG:11}). As a result in \cite{BG:11} the complete asymptotic expansion
of the partition function and all the correlators were constructed. The essential disadvantage of this method is that
it is not applicable to the multi-cut case.
 A method which allows to factorize $Q_{n,\beta}[V]$ in the multi cut case to the
product of the partition functions of the one cut "effective" potentials, was proposed in \cite{S:11}.
In the present paper the same idea is used to
 study the limit of the characteristic functional $Z_{n,\beta}[h]$ and to construct the expansion
of $Q_{n,\beta}[V]$ up to  $o(1)$ terms (see Theorem \ref{t:1}). We assume
the following conditions:

\medskip \noindent \textbf{Condition C1. }\textit{$V$ is a real analytic potential  satisfying (\ref{condV}).
The support
of the equilibrium measure is}
\begin{equation}\label{sigma}
    \sigma=\bigcup_{\alpha=1}^q\sigma_\alpha,\quad \sigma_\alpha=[a_{\alpha},b_{\alpha}];
\end{equation}
\medskip \noindent \textbf{Condition C2. }\textit{ The equilibrium density $\rho$ can be represented in the form
\begin{equation}\label{rho}
    \rho(\lambda)=\frac{1}{2\pi}P(\lambda)\Im X_\sigma^{1/2}(\lambda+i0),\quad
    \inf_{\lambda\in\sigma}|P(\lambda)|>0,
\end{equation}
where
\begin{equation}\label{X}
    X_\sigma(z)= \prod_{\alpha=1}^{q}(z-a_\alpha)(z-b_\alpha),
    \end{equation}
and we choose a branch of $X_\sigma^{1/2}(z)$ such that $X_\sigma^{1/2}(z)\sim z^q$, as $z\to+\infty$.
Moreover, the function $v$ defined by (\ref{cond_rho})
attains its maximum only if $\lambda $ belongs to  $\sigma $. }

\begin{remark} It is known (see, e.g., \cite{APS:01}) that for analytic $V$
the equilibrium density $\rho$ always
 has the form (\ref{rho}) -- (\ref{X}).
 The function $P$ in
(\ref{rho}) is  analytic and  can be represented in the form
\begin{equation}\label{P}
    P(z)=\frac{1}{2\pi i}\oint_\mathcal{L}\frac{V'(z)-V'(\zeta)}{(z-\zeta) X_\sigma^{1/2}(\zeta)}d\zeta.
    \end{equation}
Hence  condition C2 means that $\rho$ has no zeros in the internal points of $\sigma$ and behaves like
square root near the edge points. This behavior of $V$ is usually called generic.
\end{remark}

The first result of the paper is the theorem which allows us to control $Z_{n,\beta}[h]$ and $\log Q_{n,\beta}[V]$
in the one cut case up to  $O(n^{-1})$ terms. The essential difference with similar results of \cite{Jo:98},
\cite{KS:10} and \cite{BG:11} is that Theorem \ref{t:2} is applicable to a non real $h$. This fact is very important
 because the proof of Theorem \ref{t:1} is based on the  application of Theorem \ref{t:2} to a non real $h$.
 Besides, since the results of  \cite{BG:11} were obtained for  real analytic $h$,  the remainder bounds found
 here cannot be used in the proof of Theorem \ref{t:1}.

\begin{theorem}\label{t:2}
Let $V$ satisfy (\ref{condV}),  the  equilibrium
density $\rho$ (see (\ref{cond_rho}))   have  the form
(\ref{rho}) with $q=1$, and $\sigma=\mathrm{supp\,}\rho=[a,b]$.
Assume also that $V$ is  analytic  in the domain $\mathbf{D}\supset \sigma_\varepsilon$, where $\sigma_\varepsilon$
is the $\varepsilon$-neighborhood of $\sigma$.
Then:

\noindent(i) For any real $h$ with $||h^{(6)}||_\infty,||h'||_\infty\le n^{1/2}\log n$  the characteristic functional
$Z_{n,\beta}[h]$ has the form
\begin{align}\label{t2.Z}
&Z_{n,\beta}[h]=\exp\Big\{\frac{\beta}{2}\Big(\big(\frac{2}{\beta}-1\big)
(h,\nu)+\frac{1}{4}(\overline Dh,h)\Big)+n^{-1}O\big(||h'||_\infty^3+||h^{(6)}||_\infty ^3\big)\Big\},
\end{align}
where  the operator $\overline D_\sigma$ is defined by
\begin{align}&\overline D_\sigma=\frac{1}{2}(D_\sigma+D^*_\sigma),\quad D_\sigma h(\lambda)=
\frac{X^{-1/2}(\lambda)}{\pi^2 }\int_{\sigma}
\frac{h'(\mu)X^{1/2}(\mu) d\mu}{(\lambda-\mu)},\label{bar_D}\end{align}
and a non positive measure $\nu$ has the form
\begin{align}\label{t2.nu}
(\nu,h):=&\frac{1}{4}(h(b)+h(a))-\frac{1}{2\pi}\int_\sigma\frac{h(\lambda)d\lambda}{ X^{1/2}(\lambda)}-
\frac{1}{2}(D_\sigma\log P,h)
\end{align}
with $P$ defined by (\ref{P}) and $X^{1/2}(\lambda):=\Im X^{1/2}(\lambda+i0)$ with $X$ of (\ref{X}). Here and below
$||h||_\infty=\sup_{\lambda\in\sigma_\varepsilon}|h(\lambda)|$.

\noindent(ii) If $h$ is non real and  $|\beta(D_\sigma h,h)|\le \kappa\log n$ with  some absolute  $\kappa$
and $||h^{(6)}||_\infty\le n^{1/6}$, then
\begin{align}\label{t2.Za}
&Z_{n,\beta}[h]=\exp\Big\{\frac{\beta}{2}\Big(\big(\frac{2}{\beta}-1\big)
(h,\nu)+\frac{1}{4}(\overline D_\sigma h,h)\Big)\}\Big\}\Big(1+n^{-1/2}O\big(||h'||_\infty^3+||h^{(6)}||_\infty^3\big)\Big).
\end{align}

\noindent(iii) For $h=0$
\begin{align}\label{log}
\log( Q_{n,\beta}/n!)=&\frac{\beta n^2}{2}\mathcal{E}[ V]+F_\beta(n)
+n\Big(\frac{\beta}{2}-1\Big)\big((\log \rho,\rho)-1-\log 2\pi\big)\\&+r_\beta[ \rho]+O(n^{-1}),
\notag
\end{align}
 where $r_\beta[ \rho]$ is given by the integral representation of (\ref{r[V]}) and $F_\beta(n)$ corresponds
 to the linear, logarithmic  and zero order terms of the expansion in $n$   of $\log Q_{n,\beta}[V^*]$
 for $V^*(\lambda)=\lambda^2/2$. According to \cite{Fo:10}
\begin{align}\label{as_GE}
    F_\beta(n)=&n\Big(\frac{\beta}{2}-1\Big)
    \Big(\log\frac{ n\beta}{2}-\frac{1}{2}\Big)+n\log\frac{\sqrt{2\pi}}{\Gamma(\beta/2)}-c_\beta\log n+c^{(1)}_\beta,
\end{align}
where $c_\beta=\frac{\beta}{24}-\frac{1}{4}+\frac{1}{6\beta}$ and $c^{(1)}_\beta$ is some constant/ depending only on $\beta$
(for $\beta=2$, $c^{(1)}_\beta=\zeta'(1)$).
\end{theorem}
\begin{remark} Let us note that the operator $D_\sigma$ is "almost" $(-\mathcal{L}_\sigma)^{-1}$, where
$\mathcal{L}_\sigma$ is the integral operator defined by (\ref{L[,]}) for the interval $\sigma$. More precisely, if
we denote $X_{\sigma}^{-1/2}=\mathbf{1}_{\sigma}|X_{\sigma}^{-1/2}|$ with $X_{\sigma}$ of  (\ref{X})
\begin{align}
\label{DL}
D_{\sigma}\mathcal{L}_{\sigma}v=
-v+\pi^{-1}(v,\mathbf{1}_{\sigma})X^{-1/2}_{\sigma},\quad
\mathcal{L}_{\sigma}D_{\sigma} v=
-v+\pi^{-1}(v,X^{-1/2}_{\sigma})\mathbf{1}_{\sigma},\\
\mathcal{L}_{\sigma}D^*_{\sigma} v=-v+\pi^{-1}(v,X^{-1/2}_{\sigma})
\mathbf{1}_{\sigma},\quad\Rightarrow\quad
\mathcal{L}_{\sigma}\bar D_{\sigma} v=-v+\pi^{-1}(v,X^{-1/2}_{\sigma})
\mathbf{1}_{\sigma}.
\notag
\end{align}
\end{remark}
\begin{remark} For $\beta=2$  in the one-cut case we have
\begin{equation}\label{b=2}
 \log( Q_{n,2}/n!)=n^2\mathcal{E}[ V]+\frac{n}{2}\log 2\pi-\frac{1}{12}\log n+\zeta'(1)
-\frac{2}{3(b-a)^2}\log \frac{P(a)P(b)}{P_0^2}+O(n^{-1}),
\end{equation}
 where $ P_0=16/(b-a)^2$ corresponds to the Gaussian potential $V_0(\lambda)=2(2\lambda-a-b)^2/(b-a)^2$, such that the support of its equilibrium measure is $[a,b]$.
\end{remark}

 Consider the  space
\begin{equation}\label{cal-H} \mathcal{H}=\oplus_{\alpha=1}^q L_1[\sigma_\alpha].\end{equation}
 Note that we need $\mathcal{H}$ mainly as a set of functions, its topology is not important
 below.
  Define the operator $\mathcal{L}$ as
\begin{equation}\label{cal-L}
    \mathcal{L}f=\mathbf{1}_\sigma L[f\mathbf{1}_\sigma],\quad\widehat{\mathcal{L}}_\alpha f:=
    \mathbf{1}_{\sigma_\alpha} L[f\mathbf{1}_{\sigma_\alpha}].
\end{equation}
Moreover, we will consider the block diagonal operators
\begin{equation}\label{D,L}
   \overline D:=\oplus_{\alpha=1}^q \overline D_\alpha,\quad \widehat{\mathcal{L}}:=
   \oplus_{\alpha=1}^q\widehat{\mathcal{L}}_\alpha,
\end{equation}
where $\overline D_\alpha$ is defined by (\ref{bar_D}) for $\sigma_\alpha$. Introduce also
\begin{equation}\label{ti-L,G}
\widetilde{\mathcal{L}}:=\mathcal{L}-\widehat{\mathcal{L}},\quad \mathcal{G}:=(1-\overline D\widetilde{\mathcal{L}})^{-1}.
\end{equation}
An important role below belongs to a positive definite matrix of the form
\begin{equation}\label{Q}
\mathcal{Q}=\{\mathcal{Q}_{\alpha\alpha'}\}_{\alpha,\alpha'=1}^{q},\quad
\mathcal{Q}_{\alpha\alpha'}=-(\mathcal{L}\psi^{(\alpha)},\psi^{(\alpha')}),\end{equation}
where $\psi^{(\alpha)}(\lambda)=
p_\alpha(\lambda) X^{-1}(\lambda)\mathbf{1}_\sigma$ ($p_\alpha$ is a polynomial of degree $q-1$)
is a unique solution of the system of equations
\begin{equation}\label{cond_psi}
-(\mathcal{L}\psi^{(\alpha)})_{\alpha'}=\delta_{\alpha\alpha'},\quad \alpha'=1,\dots,q.
\end{equation}
Denote also
\begin{equation}\label{I[h]}I[h]=(I_1[h],\dots,I_q[h]),\quad I_\alpha[h]:=
\sum_{\alpha'}\mathcal{Q}^{-1}_{\alpha\alpha'}(h,\psi^{(\alpha')}).\end{equation}
\begin{equation}\label{mu^*}
    \mu_\alpha=\int_{\sigma_\alpha}\rho_\alpha(\lambda)d\lambda,\quad \rho_\alpha:=\mathbf{1}_{\sigma_\alpha}\rho.
\end{equation}

The main result of the paper is the following theorem:
\begin{theorem}\label{t:1} Let the potential $V$ satisfy conditions C1-C2, and let $||h^{(6)}||_\infty<\infty$. Then
\begin{align}\label{Z-mult}
    &Z_{n,\beta}[h]=
e^{\frac{\beta}{8}(\mathcal{G}\overline Dh,h)+
\Big(\frac{\beta}{2}-1\Big)(\mathcal{G}\nu,h)}\frac{\Theta(\bar I[h];\{n\bar\mu\})}
{\Theta(0;\{n\bar\mu\})}\big(1+O\big(n^{-\kappa}(||h'||_\infty^3+||h^{(6)}||_\infty ^3)\big)\big),\end{align}
where
\begin{align}\label{theta}
&\Theta( I[h];\{n\bar\mu\}):=\sum_{n_1+\dots+n_q=n_0} \exp\Big\{-\frac{\beta}{2}\Big(\mathcal{Q}^{-1}\Delta\bar n,\Delta\bar n\Big)
+\frac{\beta}{2}( \Delta\bar n,I[h])\\&
\qquad\qquad\qquad\qquad+\big(\frac{\beta}{2}-1\big)( \Delta\bar n, I[\log\overline\rho])\Big\},\notag\\
&\{n\bar\mu\}=(\{n\mu_1\},\dots,\{n\mu_q\}),\quad(\Delta\bar  n)_\alpha=n_\alpha-\{n\mu_\alpha\},
\quad n_0=\sum_{\alpha=1}^q\{n\mu_\alpha\},
\notag\end{align}
with a positive definite matrix $\mathcal{Q}$ of (\ref{Q}),  $I[h]$ defined by (\ref{I[h]}),
 and  $\log\overline\rho=(\log\rho_1,\dots,\log\rho_q)$.

For $h=0$ we have
\begin{align}\label{as_Q}
Q_{n,\beta}[V]=&\mathcal{Z}_{n,\beta}\frac{\exp\Big\{\dfrac{2}{\beta}\Big(\dfrac{\beta}{2}-1\Big)^2
\big(\widetilde{\mathcal{L}}\mathcal{G}\nu,\nu\big)\Big\}}
{\det^{1/2}(1-\overline D\widetilde{\mathcal{L}})}\Theta(0;\{n\bar\mu\})(1+O(n^{-\kappa})),\\
\mathcal{Z}_{n,\beta}[V]=&\exp\Big\{\frac{n^2\beta}{2}\mathcal{E}[V]+F_\beta(n)+n(\frac{\beta}{2}-1)
\big((\log\rho,\rho)-1-\log 2\pi\big)
\notag\\&-c_\beta(q-1)\log n
+\sum_{\alpha=1}^q(r_{\beta}[\mu_\alpha^{-1}\rho_\alpha]-c_\beta\log\mu_\alpha)\Big\},
\notag\end{align}
where $\mu_\alpha,\,\rho_\alpha$ are defined in (\ref{mu^*}),  $r_{\beta}[\rho]$ is defined in (\ref{r[V]}),
$F_\beta(n)$ and $c_\beta$ are defined in (\ref{as_GE})
 and $\det$ means the Fredholm determinant of
$\overline D\widetilde{\mathcal{L}}$ on $\sigma$.
\end{theorem}
It is evident that Theorem \ref{t:2} yields that the fluctuations of $\mathcal{N}_n[h]$ for generic $h$ are non Gaussian.
They are  Gaussian,  if there exists some $c$ such that
\begin{equation}\label{cond_G}
I_\alpha[h]=c,\quad\alpha=1,\dots,q;\quad\Leftrightarrow\quad(h-c,\psi^{(\alpha)})=0,\quad \alpha=1,\dots,q.\end{equation}
Moreover, inspecting the proof of Theorem \ref{t:1}, one can see that it is proven in fact that $\log Z_{n,\beta}[th]$
is an analytic function of $t$ for some small enough $t$. Since
\[n(p^{(1)}_{n,\beta}-\rho,h)=\frac{2}{\beta}\partial_t\log Z_{n,\beta}[th]\Big|_{t=0},\quad
\mathbf{Var}_{n,\beta}\{\mathcal{N}_n[h]\}=\big(\frac{2}{\beta}\big)^2\partial_t^2\log Z_{n,\beta}[th]\Big|_{t=0},
\]
one can find $n(p^{(1)}_{n,\beta}-\rho,h)$ and $\mathbf{Var}_{n,\beta}\{\mathcal{N}_n[h]\}$, differentiating
the r.h.s.  of (\ref{Z-mult}). It is easy to see that if conditions (\ref{cond_G}) are not fulfilled, then
both expressions contain the derivatives of $\log\Theta(I(h);\{n\mu\})$, hence they are quasi periodic functions.

\smallskip

Let us note that   relations (\ref{DL}) imply
\[-\mathcal{L}\mathcal{G}\overline{D}=(1+P^{(1)}\widetilde{\mathcal{L}}\mathcal{L}^{-1})^{-1}(1-P^{(1)})=1+P^{(1)}
\widehat F,\]
where $P^{(1)}$ is a block-diagonal operator $P_\alpha^{(1)}v=(v,X_\alpha^{-1/2})\mathbf{1}_{\sigma_\alpha}$ and
$\widehat F$ is some operator. Hence
\[(\mathcal{G}\overline{D}h)(\lambda)=-(\mathcal{L}^{-1}h)(\lambda)+\sum c_\alpha(h)\psi^{(\alpha)}(\lambda),\]
where $c_\alpha(v)$ are some constants and $\psi^{(\alpha)}$ are defined by (\ref{cond_psi}).
Besides, evidently $\mathcal{G}\overline{D}\mathbf{1}_{\sigma_\alpha}=0$, and therefore
\[0=(\mathcal{G}\overline{D}h,\mathbf{1}_{\sigma_\alpha})=-(\mathcal{L}^{-1}h,\mathbf{1}_{\sigma_\alpha})
+\sum_{\alpha'}Q_{\alpha\alpha'}c_\alpha(h),\quad\alpha=1,\dots,q.\]
 These conditions determine $c_\alpha(h)$ uniquely. On the other hand, if we define the operator
$\mathcal{D}_\sigma$ by the formula (\ref{bar_D})
with $X_\sigma$ from (\ref{X}) for the multi cut case, then  it has  the same form with some $\widetilde c_\alpha(h)$.
Hence
\[\mathcal{G}\overline{D}=\mathcal{D}_\sigma+\sum_{\alpha}
\psi^{(\alpha)}\otimes f^{(\alpha)},\]
where $f^{(\alpha)}$ are some functions of the form $X^{-1/2}_\sigma p_\alpha$ with some polynomials $p_\alpha$.

\smallskip

The paper is organized as follows. The proofs of Theorem \ref{t:2} and Theorem \ref{t:1} are given
in Section \ref{s:2} and Section \ref{s:3} respectively.
Proofs of some auxiliary
  results, used in the proof of Theorem \ref{t:1}, are given in Section \ref{s:4}.

\section{Proof of Theorem \ref{t:2} }\label{s:2}

To prove Theorem \ref{t:2} we study the Stieltjes transform
\begin{equation}\label{St_tr}
    g_{n,\beta,h}(z)=\int\frac{p_{n,\beta,h}^{(1)}(\lambda)d\lambda}{z-\lambda}
\end{equation}
 of the first marginal density
$p_{n,\beta,h}^{(1)}$ defined by (\ref{p_nl}) for $V$ replaced by $V-\frac{1}{n}h$.
Let us represent
\[g_{n,\beta,h}=g+n^{-1}u_{n,\beta,h},\]
where $g$ is the Stieltjes transform of the equilibrium density $\rho$.
 According to  \cite{S:11},
 \begin{equation}\label{eq_10}
u_{n,\beta,h}(z)= (\mathcal{K}F)(z),
\end{equation}
where the
operator $\mathcal{K}:\mathrm{Hol}[\mathbf{D}\setminus\sigma_\varepsilon]\to
\mathrm{Hol}[\mathbf{D}\setminus\sigma_\varepsilon]$ is defined by the formula
\begin{equation}\label{op_P}
(\mathcal{K}f)(z):=\frac{1}{2\pi iX^{1/2}(z)}\oint_{\mathcal{L}
}\frac{f(\zeta)d\zeta}{P(\zeta)(z-\zeta)},
\end{equation}
with the contour  $\mathcal{L}$  which
 does not contain $z$ and zeros of $P$, and
\begin{align}\label{F}
F(z)=& \int\frac{h'(\lambda)p^{(1)}_{n,\beta,h}(\lambda)}{z-\lambda}d\lambda
-\bigg(\frac{2}{\beta }-1\bigg)g'(z)\\
&-\frac{2/\beta -1}{n}u_{n,\beta,h}'(z)+\frac{1}{n}u_{n,\beta,\eta}^2(z)+\frac{1}{n}\delta_{n,\beta,h}(z),
\notag\end{align}
with
\begin{align}\label{delta}
    \delta_{n,\beta,h}(z)=\int \frac{n(n-1)p^{(2)}_{n,\beta,h}(\lambda,\mu)-n^2p^{(1)}_{n,\beta,h}(\lambda)
p^{(1)}_{n,\beta,h}(\mu)+n\delta(\lambda-\mu)p^{(1)}_{n,\beta,h}(\lambda)}{(z-\lambda)(z-\mu)}d\lambda d\mu.
\end{align}
Moreover, according to \cite{S:11} $u_{n,\beta,h}$ and $\delta_{n,\beta,h}$   satisfy the bounds:
\begin{align}\label{b_u}
&|u_{n,\beta,h}(z)|\le C_0\frac{\log n}{d^{5/2}(z)}(1+||h'||_{\infty}), \quad
|\delta_{n,\beta,h}(z)|\le C(1+||h'||_{\infty})^2\frac{\log^2 n}{d^{5}(z)},
\end{align}
if
$d(z):=\mathrm{dist}\{z,\sigma_\varepsilon\}\ge n^{-1/3}\log n$. In addition,
\begin{equation}\label{(h,p)}
 |(p_{n,\beta,h}-\rho,\varphi)|\le Cn^{-1}(||\varphi'''||_\infty+||\varphi'||_\infty).
\end{equation}
Using (\ref{b_u})  in (\ref{eq_10})
we get for $ d(z)> n^{-1/3}\log n$
\begin{align}\label{t2.a}
u_{n,\beta,h}(z)=&(\mathcal{K}\widehat h)(z)-\bigg(\frac{2}{\beta }-1\bigg)(\mathcal{K}g')(z)
\\&+n^{-1}\Big((1+||h'||_\infty^2)O(d^{-11/2}(z))+||h^{(4)}||_\infty O(d^{-3/2}(z))\Big),
\notag\end{align}
where
\[\widehat h(z):=\int\frac{h'(\lambda)\rho(\lambda)}{z-\lambda}d\lambda.\]
We note here that although (\ref{t2.a}) was obtained for $z$ inside the domain $\mathbf{D}_2$ where $V$ is an analytic
function and which does not contain zeros of $P$, we can extend (\ref{t2.a}) to
$z\not\in\mathbf{D}_2$, using that $u_{n,\beta,h}(z)$
 is analytic everywhere in $\mathbb{C}\setminus\sigma_\varepsilon$ and  behaves
like $|u_{n,\beta,h}(z)|\sim nz^{-2}$, as $z\to\infty$. Applying the Cauchy theorem, we have for any $z\not\in\mathbf{D}_2$
\begin{equation*}
u_{n,\beta,h}(z)=\frac{1}{2\pi i}\oint_{L}\frac{u_{n,\beta,h}(\zeta)d\zeta}{z-\zeta}
\end{equation*}
with the contour $L\subset\mathbf{D}_2$.

 Let us transform
 \begin{align}\label{Kh}
   (\mathcal{K}\widehat h)(z)=&\frac{1}{(2\pi i)^2 X^{1/2}(z)}\oint_{\mathcal{L}}\frac{d\zeta}{(z-\zeta)P(\zeta)}
   \int\frac{h'(\lambda)P(\lambda)|X^{1/2}(\lambda)|}{\zeta-\lambda}d\lambda\\=&
  \frac{ X^{-1/2}(z)}{2\pi}\int\frac{h'(\lambda)|X^{1/2}(\lambda)|}{z-\lambda}d\lambda.
\notag\end{align}
Similarly, we have
\begin{align}\label{Kg'}
-(\mathcal{K}g')(z)=&
\frac{1}{2\pi iX^{1/2}(z)}\oint_{\mathcal{L}}\int \frac{\rho(\lambda)d\zeta d\lambda}{(\zeta-\lambda)^2P(\zeta)(z-\zeta)}
\\=&
-\frac{X^{-1/2}(z)}{2\pi }\int_{\mathcal{L}}\frac{(\log P(\lambda))'X^{1/2}(\lambda)d\lambda}{(z-\lambda)}-
\frac{1}{2}\frac{z-c}{X(z)}+\frac{1}{2X^{1/2}(z)}.
\notag\end{align}
Hence, we obtain that for $\varphi_z(\lambda)=(z-\lambda)^{-1}$ with $d(z)\ge n^{-1/3}\log n$
\begin{align}\label{(phi,p)}
    n(p_{n,\beta,h}-\rho, \varphi_z)=&(\frac{2}{\beta}-1)(\nu,\varphi_z)+\frac{1}{2}(Dh,\varphi_z)
    \\&+n^{-1}\Big((1+||h'||_\infty^2)O(d^{-11/2}(z))+||h^{(4)}||_\infty O(d^{-3/2}(z))\Big),
\notag\end{align}
where $\nu$ is defined by (\ref{t2.nu}).

To extend (\ref{(phi,p)}) on the differentiable $\varphi$,
 consider the Poisson kernel
\begin{equation*}
\mathcal{P}_y(\lambda)=\frac{y}{\pi(y^2+\lambda^2)}.
\end{equation*}
It is easy to see that for any integrable $\varphi$
\[(\mathcal{P}_y*\varphi)(\lambda)=\frac{1}{\pi}\Im\int\frac{\varphi(\mu)d\mu}{\mu-(\lambda+iy)}.\]
Hence   (\ref{(phi,p)}) implies
\begin{align}\label{P*u}
&||\mathcal{P}_y*\nu_{n,\beta,h}||_2^2\le Cn^{-1}\Big(1+||h'||^4_\infty)y^{-11}+||h^{(4)}||^2_\infty y^{-3}\Big),
\quad |y|\ge n^{-1/3}\log n, \\
&\nu_{n,\beta,h}(\lambda):=n\Big(p^{(n)}_{\beta,h}(\lambda)-\rho(\lambda)\Big)-(\frac{2}{\beta}-1)\nu(\lambda)
-\frac{1}{2}Dh(\lambda),\notag
\notag\end{align}
where $||.||_2$ is the standard norm in $L_2(\mathbb{R})$ and the  sign measure $\nu$ is defined in (\ref{t2.nu}).

Then we use the following formula (see \cite{Jo:98}) valid for any sign measure $\nu$
\begin{equation}\label{Joh}
\int_0^\infty e^{-y}y^{2s-1}||\mathcal{P}_y*\nu_{n,\beta,h}||_2^2dy=\Gamma(2s)
\int_{\mathbb{R}}(1+2|\xi|)^{-2s}|\widehat{\nu}_{n,\beta,h}(\xi)|^2d\xi.
\end{equation}
This formula for $s=6$, the Parseval equation for the Fourier integral, and  the Schwarz inequality
yield
\begin{align*}
&\hspace{-2cm}\int_{\mathbb{R}}\varphi(\lambda)\nu_{n,\beta,h}(\lambda)d\lambda=\frac{1}{2\pi}
\int_{\mathbb{R}}\widehat{\varphi}(\xi)\widehat{\nu}_{n,\beta,h}(\xi)d\xi\\
\le&\frac{1}{2\pi}\left(\int_{\mathbb{R}}|\widehat{\varphi}(\xi)|^2(1+2|\xi|)^{2s}d\xi\right)^{1/2}
\left(\int_{\mathbb{R}}|\widehat{\nu}_{n,\beta,h}(\xi)|^2(1+2|\xi|)^{-2s}d\xi\right)^{1/2}\\
\le& \frac{C((||\varphi||_{2}+||\varphi^{(6)}||_{2})}{\Gamma^{1/2}(2s)}\left(\int_0^\infty
e^{-y}y^{2s-1}||P_y*\nu_{n,\beta,h}||_2^2dy\right)^{1/2}\\ \le &Cn^{-1}((||\varphi||_{2}+||\varphi^{(6)}||_{2})
(1+||h'||_\infty^2+||h^{(4)}||_\infty).
\end{align*}
To estimate the last integral here, we split it into two parts $|y|\ge n^{-1/3}\log n$ and $|y|< n^{-1/3}\log
n$. For the first integral we use  (\ref{P*u}) and for the second - the bound (see \cite{PS:11})
\begin{align*}
    &|u_{n,h}(z)|\le\frac{C^* n^{1/2}\log^{1/2} n}{d(z)}, \end{align*}
where $C^*$ is an $n,\eta$-independent constant which depends on $||V'+\frac{1}{n}h'||_\infty$,
$\varepsilon$, and $|b-a|$. Thus we get that for any
function $\varphi$ with bounded sixth derivative
\begin{align}\label{(u,p_1)}
&n(p_{n,\beta,h}^{(1)}-\rho, \varphi)=(\frac{2}{\beta}-1)(\nu,\varphi_z)+(Dh,\varphi_z)\\&+ (||\varphi||_{2}+||\varphi^{(6)}||_{2})
(1+||h'||_\infty^2+||h^{(4)}||_{\infty})O(n^{-1} ).
\notag\end{align}
Since
\[\frac{d}{dt}\log Z_{n,\beta}[th]=\int_{\sigma_\varepsilon}h(\lambda)p^{(1)}_{n,\beta,th}(\lambda),\]
integrating (\ref{(u,p_1)})  with $\varphi=th$ with respect to $t$, we get (\ref{t2.Z}) for real $h$.
To extend this relation to the
complex valued $h$ we use the following lemma.

\begin{lemma}\label{l:3}
Let $\{X_n\}_{n\ge 1}$ be a sequence of random variables such that
\begin{equation}\label{l3.1}
    \mathbf{E}\{e^{tX_n}\}= e^{t^2/2}(1+O(n^{-1}\log^{3/2}n)),\quad -\log^{1/2} n\le t\le\log^{1/2} n.
\end{equation}
Then the relation
\begin{equation}\label{l3.2}
    \mathbf{E}\{e^{tX_n}\}= e^{t^2/2}(1+O(n^{-1/2}\log^{3/2} n)),
\end{equation}
holds in the circle $\frac{1}{7}D$, where  $D=\{t:|t|\le \log^{1/2}n\}$
\end{lemma}
\textit{Proof.}
Consider a strip $S=\{t:|\Re t|\le\log^{1/2}n\}$. It is evident that $\mathbf{E}\{e^{tX_n}\}$
is analytic in $S$ and bounded by $2\sqrt n$ for sufficiently big $n$.
Introduce the
analytic function
\[f_n(t) := c(e^{-t^2/2}\mathbf{E}\{e^{tX_n}\}-1)n/ \log^{3/2} n,\quad  t \in D,\]
 where we choose the constant $c > 0$ such that
\[|f_n(t)|\le 1,\quad t\in \gamma=[-\log^{1/2} n,\log^{1/2} n].\]
It is possible by (\ref{l3.1}). Moreover, $f_n(t)\le n^2$, $t\in D$.
 Then, by the theorem on two constants (see \cite{Evgr}), we conclude that
\[\log |f_n(t)|\le 2(1-\omega(t;\gamma,D'))\log n,\]
where $\omega(t;\gamma,D')$ is the harmonic measure of the set $\gamma$
 with respect
to the domain $D'$ at the point $t\in D'$, where $D' := D \cap \mathbb{C}_+$. It is
well-known (see again \cite{Evgr}) that
\[\omega(t;\gamma,D')=1-\frac{2}{\pi}\Im\log\frac{1+t\log^{-1/2} n}{1-t\log^{-1/2} n}.\]
Hence $1-\omega(t;\gamma,D')\le 14\Im t/(3\pi\log^{1/2}n)$
 for $t\in\frac{1}{ 7} D'$,  and we obtain from the above inequalities that
\[
\log |f_n(t)| \le \frac{28 \log^{1/2} n}{3\pi}\Im t,\quad t\in \frac{1}{7}D'.\]
We finally deduce from the last bound that
\[\log |f_n(t)|\le \frac{1}{2}\log n,\quad\Rightarrow\quad |f_n(t)|\le n^{1/2},\,\,\,t\in\frac{1}{7}D',\]
and from the definition of $f_n$ we obtain (\ref{l3.2}).

$\square$

(iii) To prove (\ref{log}) we need to control $u_{n,\beta,h}$ up to the order $n^{-1}$.
It follows from (\ref{eq_10}) and
(\ref{F}) that for this aim we need to control zero order term of $u_{n,\beta,h}$ (which is known already) and zero order term of
$\delta_{n,\beta,h}(z)$.
It is easy to see that if  we replace $h(\lambda)$ by
$h_t^{(\alpha)}=h(\lambda)+th_{z_0}(\lambda)$  with $h_{z_0}(\lambda)=(\lambda-z_0)^{-1}$,  then
\[\delta_{n,\beta,h}(z_0)=\partial_t u_{n,\beta,h_t}(z_0)\Big|_{t=0}.\]
 It was proven in \cite{Jo:98}
that $u_{n,\beta,h_t}(z)$
is an analytic function of $t$ for small enough $t$. Hence,
integrating with respect to $t$ over the circle $|t|=C_0d^2(z_0)/2$, we get that for any $||h'||\le C_0/2$
\[\partial_tu_{n,\beta,h_t}(z)\Big|_{t=0}=
\frac{1}{\pi X_{\eta}^{1/2}(z)}\oint_{\mathcal{L}_d}
\frac{h_{z_0}'(\lambda)|X^{1/2}(\lambda)| d\lambda}{(z-\lambda)}
+n^{-1}O(d^{-11/2}(z)d^{-2}(z_0)).\]
Thus we obtain for $h=0$ and any real analytic $V$, satisfying conditions C1-C2:
\begin{equation}\label{t2.d}
\delta_{n,\beta}(z)=\frac{1}{\pi X^{1/2}(z)}\int_{\sigma}\frac{X^{1/2}(\lambda)}{(\lambda-z)^3}d\lambda
+n^{-1}O(d^{-15/2}(z))=\frac{1}{X^{2}(z)}+n^{-1}O(d^{-15/2}(z)).
\end{equation}
 Set
\begin{align*}
&V^{(0)}(z)=2(z-c)^2/d^2,\quad
c=(a+b)/2,\quad
d=(b-a)/2,\quad P_0=4/d^2,\\
&g_t(z)=tg(z)+\frac{2(1-t)}{d^2}(z-c-X^{1/2}(z)),\quad P_t(\lambda)=P_0+t(P(\lambda)-P_0),
\end{align*}
and consider  the functions $V_{t}$ of the form
\begin{equation}\label{V_t}
    V_t(\lambda)=V^{(0)}(\lambda)+t\Delta V(\lambda),\quad \Delta V(\lambda)= V(\lambda)-V^{(0)}(\lambda).
\end{equation}

Let $Q_{n,\beta}(t):=Q_{n,\beta}[V_{t}]$ be defined by (\ref{Q})
with $V$ replaced by $V_{t}$. Then, evidently, $Q_{n,\beta}(1)=Q_{n,\beta}[ V]$, and $Q_{n,\beta}(0)$
 corresponds to  $V^{(0)}$. Hence
\begin{align}\label{d_log_Q}
\frac{1}{n^2}\log Q_{n,\beta}(1)-\frac{1}{n^2}\log Q_{n,\beta}(0)&=
\frac{1}{n^2}\int_0^1dt\frac{d}{dt}\log Q_{n,\beta}(t)\\
&=-\frac{\beta}{2}\int_0^1dt\int d\lambda\Delta V(\lambda)p^{(1)}_{n,\beta}(\lambda;t),\notag
\end{align}
where $p^{(1)}_{n,\beta}(\lambda;t)$ is the first marginal density corresponding to $V_{t}$.
Using (\ref{cond_rho}), one can check that  for the distribution
(\ref{p(la)}) with $V$ replaced by $V_t$ the equilibrium density $\rho_t$ has the form
\begin{equation}\label{rho_t}
\rho_t(\lambda)=t\rho(\lambda)+(1-t)\rho^{(0)}(\lambda),\quad
\rho^{(0)}(\lambda)=\frac{2X^{1/2}(\lambda)}{\pi d^2},\quad \Delta\rho(\lambda)=\rho(\lambda)-\rho_0(\lambda)
\end{equation}
with $X, d$ of (\ref{log}).
Using (\ref{eq_10}), (\ref{F}), (\ref{t2.a}), and (\ref{t2.d}), one can write:
\begin{equation}\label{t2.1}
g_{n}(z,t)=g(z,t)+
n^{-1}u_{\beta}^{(0)}(z,t)+n^{-2}u_{\beta}^{(1)}(z,t)+O(n^{-3}),
\end{equation}
where
\begin{align}& u^{(0)}_\beta(z,t)=-\Big(\frac{2}{\beta}-1)(\mathcal{K}_tg'_t)(z),\notag\\
&u^{(1)}_\beta(z,t)=\mathcal{K}_t\Big((u_\beta^{(0)})^2-(2/\beta-1)\partial_z u_\beta^{(0)}+
\frac{1}{X^{2}}\Big)(z,t),
\label{u^1}\end{align}
and the operator $\mathcal{K}_t$ is defined by (\ref{op_P}) with $P$ replaced by $P_t=P_0+(1-t)P$.

Substituting (\ref{t2.1}) in  the last integral in (\ref{d_log_Q}), we get
\begin{align}\label{t2.ii}
\log Q_{n,\beta}[V]=&\log Q_{n,\beta}[V^{(0)}]
-n^2\frac{\beta}{2}\mathcal{E}[V^{(0)}]+n^2\frac{\beta}{2}\mathcal{E}[ V]\\
&+n(\frac{\beta }{2}-1)\int_0^1dt(\Delta V(\lambda),\nu_t)-\int_0^1dt\frac{\beta}{4\pi i}\oint\Delta V(z)
u^{(1)}_{\beta}(z,t)dz+O(n^{-1}).
\notag\end{align}
Write $\Delta V=2L[\Delta \rho]+v^{(0)}$ where $v^{(0)}$ is a  constant from (\ref{cond_psi}), corresponding
to $V^{(0)}$ (recall that we assumed that corresponding  constant for $V$, is zero).
Then, taking into account (\ref{t2.nu}),
we get
\begin{align*}
(\Delta V(\lambda),\nu)=\frac{1}{4}(\Delta V(a)+\Delta V(b))-\frac{v^{(0)}}{2}
-(L[\Delta \rho],D\log P_t).
\end{align*}
Then (\ref{DL}) yields
\begin{align*}
(L[\Delta \rho],D\log P_t)=(\Delta \rho,LD\log P_t)=-( \Delta\rho,\log
P_t).
\end{align*}
Now we can integrate with respect to $t$ and obtain
\begin{align}\label{(V,nu)}
&\int_0^1dt(\Delta V(\lambda),\nu)=\frac{1}{4}(\Delta V(a)+\Delta V(b))-\frac{v^{(0)}}{2}\\
&+\int_{\sigma}\rho(\lambda)\log P(\lambda)d\lambda-\int_{\sigma}\rho_0(\lambda)\log P_0(\lambda)d\lambda\notag\\
&=\int_{\sigma}\rho(\lambda)\log \rho(\lambda)d\lambda-1-\log 2\pi
+\log (d/2),
\notag\end{align}
since
\begin{align*}
&\int_{\sigma}\rho(\lambda)\log  X^{1/2}(\lambda)d\lambda-\frac{1}{4}( V(a)+ V(b))
=\frac{1}{2}(L[\rho](a)+L[\rho](b))-\frac{1}{4}( V(a)+ V(b))=0,\\
&\int_{\sigma}\rho_0(\lambda)\log P_0 d\lambda=-2\log  (d/2)
\quad V^{(0)}(a)=V^{(0)}(b)=2,\quad v^{(0)}=2\log(d/2).
\end{align*}
In addition, changing the variables in the corresponding integrals, we have
\begin{align*}
\log Q_{n,\beta}[V^{(0)}]=&\log Q_{n,\beta}^{*}+\Big(\frac{n^2\beta}{2}
+n(1-\beta/2)\Big)\log\frac{d}{2},\\
\frac{n^2\beta}{2}\mathcal{E}[V^{(0)}]=&-\frac{3n^2\beta}{8}+\frac{n^2\beta}{2}\log\frac{d}{2}.
\end{align*}
These relations combined with (\ref{t2.ii}), (\ref{(V,nu)}) and  (\ref{u^1}) imply (\ref{log}) with
\begin{align}\notag
r_\beta[ \rho]:=&-\frac{1}{2\pi i}\int_0^1dt\oint_L\Delta V(z)
u^{(1)}_{\beta}(z,t)dz\\=&\frac{1}{(2\pi )^2}\int_0^1dt\oint_Ldz\frac{\Delta V(z)}{X^{1/2}(z)}\oint_{L'}
d\zeta\frac{\Big((u^{(0)})^2-(\frac{2}{\beta}-1)\partial_\zeta u^{(0)}+
X^{-2}\Big)(\zeta,t)}{(z-\zeta)(P_0+t\Delta P(\zeta))},
\label{r[V]}\end{align}
where the contour $L$ contains $L'$, which contains $\sigma_\varepsilon$,   all zeros of $P_t$ are outside of $L$,
 and $u^{(0)}_{\beta}$ is defined in (\ref{u^1}). For $\beta=2$ $u^{(0)}_{\beta}=0$, hence we can leave
 only $X^{-2}(\zeta)$ in the last numerator and take the integral with respect
 to  $\zeta$. Taking into account that
 \[\Delta V'(z)=2\Delta g(z)+\Delta P(z)X^{1/2}(z),\]
 and $\Delta g(z)\sim Cz^{-2}$, as $z\to\infty$, we have
\begin{align*}
\frac{a-b}{2}\oint_{L}\frac{\Delta V(z) dz}{X^{1/2}(z)(z-a)}=&\oint_L\frac{\Delta V'(z)(z-b)^{1/2} dz}{(z-a)^{1/2}}\\=&
\oint_L\frac{2\Delta g(z)(z-b)^{1/2} dz}{(z-a)^{1/2}}+\oint_L\frac{\Delta P(z)X^{1/2}(z)(z-b)^{1/2} dz}{(z-a)^{1/2}}=0,\\
\oint_{L}\frac{\Delta V(z) dz}{X^{1/2}(z)(z-a)^2}=&\frac{2}{3}\oint_L\frac{\Delta V'(z) dz}{(z-b)^{1/2}(z-a)^{3/2}}-
\frac{1}{3}\oint_L\frac{\Delta V(z) dz}{(z-b)^{3/2}(z-a)^{3/2}}\\&=
\frac{2}{3}\oint_L\frac{\Delta P(z) dz}{(z-a)}=\frac{4\pi i}{3}\Delta P(a),
\end{align*}
and similar relation for  integrals with $(z-a)$ replaced by $(z-b)$. Thus we obtain (\ref{b=2}).
$\square$

\section{Proof of Theorem \ref{t:1}}\label{s:3}
Denote
\begin{align}\label{s_e}
&\sigma_\varepsilon=\bigcup_{\alpha=1}^q\sigma_{\alpha,\varepsilon},\quad
\sigma_{\alpha,\varepsilon}=[a_{\alpha}-\varepsilon,b_{\alpha}+\varepsilon],\\
&\mathrm{dist\,}\{\sigma_{\alpha,\varepsilon},\sigma_{\alpha',\varepsilon}\}
>\delta>0,\quad \alpha\not=\alpha'.
\notag\end{align}
It is known (see  \cite[Lemmas 1,3]{BPS:95} and \cite[Theorems 11.1.4, 11.1.6]{PS:11})
that if we replace in (\ref{p(la)}) and (\ref{p_nl}) the integration over $\mathbb{R}$
by the integration over $\sigma_\varepsilon$, then   the new  partition function $Q_{n,\beta}^{(\varepsilon)}[ V]$
and the old one $Q_{n,\beta}[V]$ satisfy the inequality
\begin{align*}
 \big|Q_{n,\beta}[ V]/Q_{n,\beta}^{(\varepsilon)}[ V]-1\big|\le
e^{-n\beta d_\varepsilon}
\end{align*}
 Thus, it suffices to study $Q_{n,\beta}^{(\varepsilon)}[V]$  instead of $Q_{n,\beta}[V]$. Starting from this moment,
 we assume  that the
replacement of the integration domain is made,  but we will omit superindex $\varepsilon$.

Consider the "approximating" function $H_a$ (Hamiltonian)
\begin{align}\label{H_a}
H_a(\lambda_1\dots \lambda_n)=&-n\sum  V^{(a)}(\lambda_i)+
\sum_{i\not=j}\log|\lambda_i-\lambda_j|\Big(\sum_{\alpha=1}^q\mathbf{1}_{\sigma_{\alpha,\varepsilon}}(\lambda_i)
\mathbf{1}_{\sigma_{\alpha,\varepsilon}}(\lambda_j)\Big)-n^2\Sigma^*,\\
V^{(a)}(\lambda)=&\sum_{\alpha=1}^qV^{(a)}_\alpha(\lambda),\quad
V^{(a)}_\alpha(\lambda)=\mathbf{1}_{\sigma_{\alpha,\varepsilon}}(\lambda)
\Big(V(\lambda)-2\int_{\sigma\setminus\sigma_\alpha}\log|\lambda-\mu|\rho(\mu)d\mu\Big),
\label{V^a}\end{align}
where $V^{(a)}_\alpha(\lambda)$ is the "effective potential".
It is easy to check that (\ref{cond_rho}) implies
\begin{equation}\label{V=2L}
    V^{(a)}_\alpha=2L[\rho_\alpha].
\end{equation}
The "cross energy" $\Sigma^*$ in (\ref{H_a}) has the form
\begin{align}\label{S*}\Sigma^*:=&
 \sum_{\alpha\not=\alpha'}L[\rho_\alpha,\rho_{\alpha'}].
 \end{align}
Then
\begin{eqnarray}
H(\lambda_1\dots \lambda_n)&=&H_a(\lambda_1\dots \lambda_n)+\Delta H(\lambda_1\dots \lambda_n),
\quad \lambda_1,\dots,\lambda_n\in\sigma_\varepsilon,\label{DeltaH}\\
\Delta H(\lambda_1\dots \lambda_n)&=&\sum_{i\not=j}\log|\lambda_i-\lambda_j|\sum_{\alpha\not=\alpha'}
\mathbf{1}_{\sigma_{\alpha,\varepsilon}}(\lambda_i)\mathbf{1}_{\sigma_{\alpha',\varepsilon}}(\lambda_j)-2n\sum_{j=1}^n\widetilde V(\lambda_j)+n^2\Sigma^*,\notag\\
\widetilde V(\lambda)&=&\sum_{\alpha=1}^q\mathbf{1}_{\sigma_{\alpha,\varepsilon}}(\lambda)
\int_{\sigma\setminus\sigma_\alpha}\log|\lambda-\mu|\rho(\mu)d\mu.
\notag\end{eqnarray}
Set
\begin{equation}\label{1_n}
    \overline n:=(n_1,\dots,n_q),\quad |\bar n|:=\sum_{\alpha=1}^q n_\alpha,\quad
    \mathbf{1}_{\overline n}(\overline \lambda):=\prod_{j=1}^{n_1}\mathbf{1}_{\sigma_{1,\varepsilon}}(\lambda_j)\dots
\prod_{j=|\bar n|-n_q+1}^{n}\mathbf{1}_{\sigma_{q,\varepsilon}}(\lambda_j).
\end{equation}
The key observation which explains our motivation to introduce $H_a$ and $\Delta H$ is that
\begin{align}\label{De_H}
 &\mathbf{1}_{\overline n}(\overline \lambda)\Delta H(\overline \lambda)=
\mathbf{1}_{\overline n}(\overline \lambda)
\sum_{\alpha\not=\alpha'}\sum_{j,k=1}^n\mathbf{1}_{\sigma_{\alpha,\varepsilon}}
(\lambda_j)\mathbf{1}_{\sigma_{\alpha',\varepsilon}}(\lambda_k)
\\&\cdot\int\log|\lambda-\mu|(\delta(\lambda_j-\lambda)-\frac{n}{n_\alpha}\rho(\lambda))
(\delta(\lambda_k-\mu)-\frac{n}{n_{\alpha'}}\rho(\mu))d\lambda d\mu.
\notag\end{align}
It was proven in \cite{S:11} that $\mathbf{E}_{n,\beta}\{\Delta H\}=O(1)$, hence this term is "smaller" than $H_a$.
On the other hand, by the construction, $H_a$ does not contain an "interaction" between different intervals $\sigma_\alpha$,
so it is possible to apply to it the result of Theorem \ref{t:2}. This idea was used in \cite{S:11} to prove that
$Q_{\bar n,\beta}[V]$ can be factorized into a product of one-cut partition functions corresponding to $V^{(a)}_\alpha$
with the error $O(1)$.
Here we are doing the next step.

It is easy to see that if we denote
\begin{align}\label{Q_k}
\quad Q_{\bar n,\beta}[V]=&\int_{\sigma_\varepsilon^n}\mathbf{1}_{\overline n}(\overline \lambda) e^{\beta H(\lambda_1\dots
\lambda_n)/2}d\lambda_1\dots d\lambda_n,
\end{align}
then
\begin{align}\label{Q=sum}
\frac{Q_{n,\beta}[V]}{n!}=\sum_{ |\bar n|=n}\frac{Q_{\bar n,\beta}[V]}{n_1!\dots n_q!}.\end{align}
According to the result of \cite{S:11},
\[\frac{Q_{\bar n,\beta}[V]}{n_1!\dots n_q!}\Big/\frac{Q_{n,\beta}[V]}{n!}\le e^{-c(\Delta n,\Delta n)},\quad\Delta n:=
(n_1-\mu_1n,\dots,n_q-\mu_qn),\]
where $\mu_\alpha$ were defined in (\ref{mu^*}). Hence, to construct the expansion of $Q_{n,\beta}[V]$, it is enough
to consider in (\ref{Q=sum}) only those terms for which
\begin{align}\label{b_n}(\Delta n,\Delta n)\le\log^2n.\end{align}
To manage with these terms we are going "to linearize" the quadratic form (\ref{De_H}) by using the integral Gaussian representation
(see (\ref{H-Str}) below). Then we will apply Theorem \ref{t:2} inside the integrals and then integrate the result.
As the first step in this direction one should find a good approximation of the integral quadratic form (\ref{De_H})
 by some
quadratic form of the finite rank. To this aim consider the  space of functions
\[\mathcal{H}_{\varepsilon}=\oplus_{\alpha=1}^qL_1[\sigma_{\alpha,2\varepsilon}],\]
and the operator $\mathcal{L}$ with the kernel $\log|\lambda-\mu|$. It has a block structure
$\{\mathcal{L}_{\alpha,\alpha'}\}_{\alpha,\alpha'=1}^q$. Denote $\widehat{\mathcal{L}}$ its block-diagonal part and
by $\widetilde{\mathcal{L}}$ the off diagonal part.

 Consider the Chebyshev polynomials $\{p^{(\alpha)}_{k}\}_{k=0}^\infty $ on
$\sigma_{\alpha,2\varepsilon}$ the corresponding orthonormal system of the functions
\[p^{(\alpha)}_{k}(\lambda)=\cos k\Big(\arccos\Big(\frac{2\lambda-(a_\alpha+b_\alpha)}
{b_\alpha-a_\alpha+4\varepsilon}\Big)\Big),\quad
\varphi^{(\alpha)}_{k}(\lambda)=p^{(\alpha)}_{k}(\lambda)|X^{-1/4}_{\sigma_{\alpha,2\varepsilon}}(\lambda)|.\]
 It is well known that $\{\varphi^{(\alpha)}_{k}\}_{k=0}^\infty$ make an orthonormal basis in
$L_2[\sigma_{\alpha,2\varepsilon}]$, hence we can write
\begin{align}\label{repr_L}
    &\mathbf{1}_{\sigma_{\alpha,\varepsilon}}(\lambda)\mathbf{1}_{\sigma_{\alpha',\varepsilon}}(\mu)
    \log|\lambda-\mu|
    =
    \sum_{k_\alpha,k_{\alpha'}=1}^\infty \mathcal{L}_{k,\alpha;k',\alpha'}p^{(\alpha)}_{k}(\lambda)
    p^{(\alpha')}_{k'}(\mu),\notag\\&
\mathcal{L}_{k,\alpha;k',\alpha'}=\int\int
\log|\lambda-\mu|
    \frac{p^{(\alpha)}_{k}(\lambda)}{|X^{1/2}_{\sigma_{\alpha,2\varepsilon}}(\lambda)|}
    \frac{p^{(\alpha')}_{k'}(\mu)}{|X^{1/2}_{\sigma_{\alpha',2\varepsilon}}(\mu)|}d\lambda d\mu.
\notag\end{align}
\begin{proposition}\label{p:appr} There exists  $C,d>0$ such that for all $\alpha\not=\alpha'$
\begin{equation}\label{p1.1}
|\mathcal{L}_{k,\alpha;k',\alpha'}|\le Ce^{-d(k+k')}.
\end{equation}
\end{proposition}
The proof of the proposition is given in Section \ref{s:4}.

Proposition \ref{p:appr} implies, in particular, that if  we choose $M=[\log^2 n]$, then uniformly in $\lambda,\mu$
\begin{align}\label{appr}
\mathbf{1}_{\sigma_{\alpha,\varepsilon}}(\lambda)\mathbf{1}_{\sigma_{\alpha',\varepsilon}}(\mu)
\log|\lambda-\mu|=&
   \mathbf{1}_{\sigma_{\alpha,\varepsilon}}(\lambda)\mathbf{1}_{\sigma_{\alpha',\varepsilon}}(\mu)
    \sum_{k,k'=1}^M \mathcal{L}_{k,\alpha;k',\alpha'}
  p^{(\alpha)}_{k}(\lambda)
   p^{(\alpha')}_{k'}(\mu)\\&+O(e^{-d\log^2n}).
\notag\end{align}
 Consider the  matrix
$\widetilde{ \mathcal{L}}^{(M)}:=\{\mathcal{L}_{k,\alpha;k',\alpha'}\}_{\substack{k,k'=1,\dots,M;\\
\alpha,\alpha'=1,\dots,q,\alpha\not=\alpha'}}$. It is a symmetric block matrix in which
 the  block $\{\mathcal{L}_{k,\alpha;k',\alpha'}\}_{\substack{k,k'=1,\dots,M}}$
corresponds to the kernel $\widetilde{ \mathcal{L}}^{(M)}_{\alpha\alpha'}$ which is the r.h.s. sum of (\ref{appr}).

Now we would like to represent  the matrix  $\widetilde{ \mathcal{L}}^{(M)}$
as a difference of two positive matrices. To this aim consider  the integral operator $\mathcal{A}$ in
$\mathcal{H}_{\varepsilon}$ with a
kernel $a(|\lambda-\mu|)$ of the form
\begin{align}\label{a}
    a(\lambda)=\left\{\begin{array}{ll}\log d^{-1}+a_0(\lambda/d)-a_0(1),&0\le \lambda\le d,\\
    \log |\lambda|^{-1},&d\le \lambda,
     \end{array}\right.
\end{align}
where the function
\[
a_0(\lambda)=\frac{3}{4}\lambda^4-\frac{8}{3}\lambda^3+3\lambda^2\]
is chosen in such a way that $a(\lambda)$  and its first 4 derivatives are continuous at
$\lambda=d$, and the third derivative of $a(|\lambda|)$ has a jump at $\lambda=0$.
\begin{lemma}\label{l:pos}
The integral operator $A$ with the kernel $a(|\lambda-\mu|)$ is positive in $L_2(\Delta)$ where $\Delta\subset[-1,1]$ is any finite
system of intervals in $\mathbb{R}$. Moreover, the integral operator  with a kernel $\log|\lambda-\mu|^{-1}-
a(|\lambda-\mu|)$ is positive in $L_2(\Delta)$
\end{lemma}
\begin{remark} One can easily see that if we choose $a_0(\lambda)=\lambda-1$, then the operator $A$ will be also positive,
but in this case the Fourier transform $\widehat a(k)\sim k^{-2}$, as $k\to\infty$, while we  need below
 $\widehat a(k)\sim k^{-4}$.
\end{remark}
 Let $\widehat{\mathcal{A}}$ be a block-diagonal part of $\mathcal{A}$. By the construction and
 the lemma we have
 \begin{equation}\label{pr_A}
\widetilde{\mathcal{L}}=\widehat{\mathcal{A}}-\mathcal{A},\quad \mathcal{A}\ge 0,\quad \widehat{\mathcal{A}}\ge 0,
\quad \widehat{\mathcal{A}}\le \widehat{\mathcal{L}}
\end{equation}

By (\ref{pr_A}), if we consider the matrix of $\mathcal{A}^{(M)}$ and $\widehat{ \mathcal{A}}^{(M)}$ at the same basis
we obtain
\[\mathcal{L}_{k,\alpha;k',\alpha'}= \mathcal{A}^{(M)}_{k,\alpha;k',\alpha'}-
\widehat{\mathcal{A}}^{(M)}_{k,\alpha;k',\alpha'}.\]
Since $\mathcal{A}^{(M)}$ and $\widehat{ \mathcal{A}}^{(M)}$ are positive matrices they can be written in the form
$\mathcal{A}^{(M)}=S^2$, $\widehat{ \mathcal{A}}^{(M)}=\widehat S^2$. Thus
\begin{align*}\Delta H=&\sum_{j,\alpha'}
 \Big(\sum_{l=1}^n\sum_{k,\alpha}
 (\widehat S_{j,\alpha';k,\alpha}
 \big(p^{(\alpha)}_{k}(\lambda_l)
 -c^{(\alpha)}_{k}\big)\Big)^2\\&-\sum_{j,\alpha'}\Big(\sum_{l=1}^n\sum_{k,\alpha}
 ( S_{j,\alpha';k,\alpha}
 \big(p^{(\alpha)}_{k}(\lambda_l)
 -c^{(\alpha)}_{k}\big)\Big)^2
 +O(e^{-d\log^2n}),\end{align*}
 where
 \[c^{(\alpha)}_{k}=\frac{n}{n_{\alpha}}(p^{(\alpha)}_{k},\rho^{(\alpha)}).\]
 Using the representations
\begin{equation}\label{H-Str}
e^{\beta x^2/2}=\sqrt{\frac{\beta}{2\pi}}\int du e^{\beta xu/2-\beta u^2/8},\quad
e^{-\beta x^2/2}=\sqrt{\frac{\beta}{2\pi}}\int du e^{i\beta xu/2-\beta u^2/8},\end{equation}
we obtain
\begin{align}\label{t1.1}
Q_{\bar n}[h]:=&\Big(\frac{\beta}{2\pi}\Big)^{Mq}\mathcal{Z}_{n,\beta}[V]e^{n^2\beta\Sigma^*/2}
\int\prod_{\alpha=1}^q\prod_{k=1}^Mdu^{(1)}_{k,\alpha}
du^{(2)}_{k,\alpha}
e^{-\frac{\beta}{8}( \bar u,\bar u)}I_{\bar n}(\bar u),\end{align}
where $\bar u:=(\bar u^{(1)},\bar u^{(2)})$,
\begin{align}\label{I(u)}
I_{\bar n}(\bar u):=&\mathcal{Z}_{n,\beta}^{-1}[V]e^{-n^2\beta\Sigma^*/2}
\prod\frac{Q_{ n_\alpha}[\mu_\alpha^{-1}V^{(a)}_\alpha-n_\alpha^{-1}\widetilde h_\alpha]}{n_\alpha!},\\
\widetilde h_\alpha(\lambda)=&(n_\alpha/\mu_\alpha-n)V^{(a)}_\alpha
+h_\alpha(\lambda)+s^{(\alpha)}(\bar u,\lambda)-\frac{n}{n_\alpha}\big(s^{(\alpha)}(\bar u,.),\rho_\alpha),
\notag\\
s^{(\alpha)}(\bar u,\lambda):=&\sum_{j,k,\alpha'}\Big(
\widehat S_{j,\alpha';k,\alpha}u_{j,\alpha'}^{(1)}
+iS_{j,\alpha';k,\alpha}u_{j,\alpha'}^{(2)}\Big)p^{(\alpha)}_{k}(\lambda).
\notag\end{align}

We are going to apply   (\ref{t2.Z}) to
$Q_{ n_\alpha}[\mu_\alpha^{-1}V^{(a)}_\alpha-n_\alpha^{-1}\widetilde h_\alpha]$. According to Theorem \ref{t:2},
it can be done   for those $\bar u:=(\bar u^{(1)},\bar u^{(2)})$ which provide the condition
\begin{equation}\label{U_1}
U_1=\{\bar u:=(\bar u^{(1)},\bar u^{(2)}):\sum_\alpha|(D_\alpha\widetilde h_\alpha,\widetilde h_\alpha)|\le c_0\log n\}.
\end{equation}
Remark that evidently $||\widetilde h^{(6)}_\alpha||_\infty\le CM^7=C\log^{14}n$. It
 will be proven below (see Lemma \ref{l:DL}) that the integral over  the compliment of $U_1$ gives us $O(n^{-\kappa})$,
 so we should concentrate on $\bar u\in U_1$.

For $\bar u\in U_1$ (\ref{t2.Z}) implies
\begin{align*}&\frac{Q_{ n_\alpha}[\mu_\alpha^{-1}V^{(a)}_\alpha-n_\alpha^{-1}\widetilde h_\alpha]}{n_\alpha!}
=\exp\Big\{\frac{\beta}{2}\big(\frac{n_\alpha}{\mu_\alpha}\big)^2\mathcal{E}_\alpha+F_\beta(n_\alpha)+
n_\alpha(\frac{\beta}{2}-1)\big(\big(\log\frac{\rho_\alpha}{\mu_\alpha},\frac{\rho_\alpha}{\mu_\alpha}\big)-1-\log 2\pi\big)
\\
&\hskip3cm+\frac{\beta}{2}r_\beta[\mu_\alpha^{-1}\rho_\alpha]+\frac{\beta}{2}
\Big(\widetilde h_\alpha,\frac{n_\alpha}{\mu_\alpha}\rho_{\alpha}+
(\frac{2}{\beta}-1)\nu_\alpha)\Big)+
\frac{\beta}{8}\Big(D_\alpha\widetilde h_\alpha,\widetilde h_\alpha\Big)\Big\},
\end{align*}
where  $r_\beta[\rho]$ is defined in (\ref{log}), $F_\beta(n)$  is defined in (\ref{as_GE}), and $\mathcal{E}_\alpha$
is the energy, corresponding to the potential $V^{(a)}_\alpha$ on $\sigma_\alpha$. In view of  (\ref{V=2L}) we have
\[\mathcal{E}_\alpha=L[\rho_\alpha,\rho_\alpha]-(V^{(a)}_\alpha,\rho_\alpha)=-L[\rho_\alpha,\rho_\alpha].\]
The definition of $\widetilde h_\alpha$ (\ref{I(u)}) and (\ref{V=2L}) yield
\begin{align*}
\big(\widetilde h_\alpha,\frac{n_\alpha}{\mu_\alpha}\rho_{\alpha}+(\frac{2}{\beta}-1)\nu_\alpha\big)=&
\big( h_\alpha,\frac{n_\alpha}{\mu_\alpha}\rho_{\alpha}+(\frac{2}{\beta}-1)\nu_\alpha\big)+
2\frac{n_\alpha}{\mu_\alpha}\Big(\frac{n_\alpha}{\mu_\alpha}-n\Big)L[\rho_\alpha,\rho_\alpha]\\&+
\Big(\frac{n_\alpha}{\mu_\alpha}-n\Big)(\frac{2}{\beta}-1)(V^{(a)}_\alpha,\nu_\alpha)+
 \big(s^{(\alpha)}( u),\big(\frac{n_\alpha}{\mu_\alpha}-n\big)\rho_{\alpha}+(\frac{2}{\beta}-1)\nu_\alpha\big).
\end{align*}
Moreover,   (\ref{DL}) implies that
\[DL\rho_\alpha=-\rho_\alpha+\frac{\mu_\alpha}{\pi} X_\alpha^{-1/2},\]
where here and below we denote $X_\alpha^{-1/2}=|X_{\sigma_\alpha}^{-1/2}|\mathbf{1}_{\sigma_\alpha}$ with $X_\sigma$ of (\ref{X}). Hence, using (\ref{V=2L}), we obtain
\begin{align*}
\frac{1}{4}\Big(\bar D_\alpha\widetilde h_\alpha,\widetilde h_\alpha\Big)=&
\frac{1}{4}\Big(\bar D_\alpha h_\alpha, h_\alpha\Big)+
\Big(\frac{n_\alpha}{\mu_\alpha}-n\Big)\Big(\frac{\mu_\alpha}{\pi}X_\alpha^{-1/2}-\rho_\alpha,h_\alpha\Big)
\\&+\Big(\frac{n_\alpha}{\mu_\alpha}-n\Big)^2\Big(-L[\rho_\alpha,\rho_\alpha]+
\frac{\mu_\alpha^2}{\pi^2}L[X^{-1/2}_\alpha,X^{-1/2}_\alpha] \Big)\\
& \Big(s^{(\alpha)}(u),\Big(\frac{n_\alpha}{\mu_\alpha}-n\Big)\Big(\frac{\mu_\alpha}{\pi}X_\alpha^{-1/2}-\rho_\alpha\Big)
+\frac{Dh}{2}\Big)
+\frac{1}{4}\Big(\bar D_\alpha s^{(\alpha)}(u),s^{(\alpha)}(u)\Big).
\end{align*}
In addition,
\[-\big(\nu_\alpha,\mu_\alpha^{-1}V^{(a)}_\alpha\big)+
\big(\log\frac{\rho_\alpha}{\mu_\alpha},\frac{\rho_\alpha}{\mu_\alpha}\big)
=\frac{1}{\pi}(\log \frac{\rho_\alpha}{\mu_\alpha},X^{-1/2}_\alpha).\]
Hence, if we introduce the notations
\begin{align}\notag
    &X^{-1/2}_{\bar n}=\pi^{-1}\big((n_1-n\mu_1)X^{-1/2}_1,\dots,(n_q-n\mu_q)X^{-1/2}_q\big),\\&
    s(u)=(s^{(1)}(u),\dots,s^{(q)}(u)),\notag\\
     &h=(h_1,\dots,h_q),\quad \nu=(\nu_1,\dots,\nu_q),\quad T=\big(\log \frac{\rho_1}{\mu_1},\dots,\log \frac{\rho_q}{\mu_q}\big),
\label{de_rho}\end{align}
then for $\bar u\in U_1$ we obtain finally
\begin{align}\label{I_2}
I_{\bar n}(\bar u)=&I_{\bar n}^{(0)}\cdot I_{\bar n}^{(1)}\cdot I_{\bar n}^{(2)}(u)(1+O(n^{-\kappa})),\quad\bar u\in U_1,\\
I_{\bar n}^{(0)}=&\exp\Big\{-n\big(\frac{\beta}{2}-1\big)\sum\mu_\alpha\log\mu_\alpha+\sum F_\beta(n_\alpha)-F_\beta(n)\Big\},\notag
\\I_{\bar n}^{(1)}=&\exp\Big\{\frac{\beta}{8}\big(\bar D h, h\big)+
\frac{\beta}{2}\big(\widehat L  X^{-1/2}_{\bar n}, X^{-1/2}_{\bar n}\big)\Big)+
\big(\frac{\beta}{2}-1\big)\big(T,X^{-1/2}_{\bar n}\big)
\notag\\&+\big(h,\frac{\beta}{2}X^{-1/2}_{\bar n}-
\big(\frac{\beta}{2}-1\big)\nu\big)\Big\},\notag\\I_{\bar n}^{(2)}(\bar u)=&\exp\Big\{\frac{\beta}{8}(\bar D s(u),s(u))+
\frac{\beta}{2}\Big(s(u), X^{-1/2}_{\bar n}+\frac{1}{2}\bar Dh+(\frac{2}{\beta}-1)\nu\Big)
\Big\}.
\notag\end{align}
To integrate with respect to
$\bar u$, we introduce the block
matrices
\begin{align*}
&E=\left(\begin{array}{ll}  I&  I \\ I& I \end{array}\right),\quad\mathcal{D}=
\bar D^{(M)}E=\left(\begin{array}{ll}  \bar D^{(M)}& \bar D^{(M)}  \\ \bar D^{(M)} & \bar D^{(M)} \end{array}\right)
\quad\mathcal{S}=\left(\begin{array}{ll}\widehat S& 0\\
0&iS\end{array}\right)\\&
\bar D^{(M)}_{\alpha,k;\alpha',k'}:=\delta_{\alpha,\alpha'}\big(\bar D_\alpha p^{(\alpha)}_{k},
p^{(\alpha)}_{k'}\big),\quad
\mathcal{F}=I-\mathcal{S}\mathcal{D}\mathcal{S}.
\notag\end{align*}
Thus,
\begin{align}\label{t2.3}
&\hskip-1cm e^{-\frac{\beta}{8}(\overline u,\overline u)}I^{(2)}_{\bar n}(u)=
\exp\Big\{-\frac{\beta}{8}\Big(\mathcal{F} \overline u,\overline u\Big)
+\frac{\beta}{4}\big(\mathcal{S}\bar u,\bar R^{(M)})\Big\},
\end{align}
where
\begin{equation}\label{bar_r}
\bar R^{(M)}:=(\bar r^{(M)},\bar r^{(M)})\quad\bar r^{(M)}=\{r^{(M)}_{\alpha,k}\},\quad r^{(M)}_{\alpha,k}
:=(2 X^{-1/2}_{\bar n}+2(1-\frac{2}{\beta})\nu+\bar Dh,p^{(\alpha)}_{k}\big).\end{equation}
\begin{lemma}\label{l:DL} There exist $\delta_1,\kappa_1>0$, such that
\begin{equation}\label{DL.1}
\Re(\mathcal{F}\bar u,\bar u)\ge\delta_1(\bar u,\bar u),
\end{equation}
and $I_{\bar n}$, $I^{(1)}_{\bar n}$, $I^{(2)}_{\bar n}$ defined by (\ref{I_2}) satisfy the bounds
\begin{align}\label{DL.1a}
&\Big(\frac{\beta}{2\pi}\Big)^{Mq}\int_{U_1^c} e^{-\beta(\bar u,\bar u)/8}|I_{\bar n}(\bar u)|d\bar u\le n^{-\kappa_1},\\
&\Big(\frac{\beta}{2\pi}\Big)^{Mq}\int_{U_1^c} e^{-\beta(\bar u,\bar u)/8}|I_{\bar n}^{(1)}I_{\bar n}^{(2)}(\bar u)|d\bar u\le n^{-\kappa_1},
\notag\end{align}
where  $U_1^c$ is a complement of $U_1$ from (\ref{U_1}).
\end{lemma}
The lemma and (\ref{t2.3}) imply that the integral over $\bar u$ of $I_{\bar n}(u)$ coincides with the integral
over $\bar u$ of  $I_{\bar n}^{(1)}I_{\bar n}^{(2)}(\bar u)$ up to the error $O(n^{-\kappa_1})$. By the virtue of
the standard  Gaussian integration formulas we obtain
\begin{align}\label{I_*}
I_{\bar n}^{*}:=&\Big(\frac{\beta}{2\pi}\Big)^{Mq}\int e^{-\beta(\bar u,\bar u)/8}I_2(\bar u)du\\
=&\mathrm{det}^{-1/2}\mathcal{F}\exp\Big\{
\frac{\beta}{8}\big(\mathrm{Tr}(\mathcal{S}\mathcal{F}^{-1}\mathcal{S}E)\bar r^{(M)},\bar r^{(M)}\big)\Big\}
(1+O(n^{-\kappa})),\notag\end{align}
where $\bar r^{(M)}$ is defined in (\ref{bar_r}) and
\begin{align*}
\mathrm{Tr}(\mathcal{S}\mathcal{F}^{-1}\mathcal{S}E)=(\mathcal{S}\mathcal{F}^{-1}\mathcal{S}E)_{11}+
(\mathcal{S}\mathcal{F}^{-1}\mathcal{S}E)_{22}.\notag\end{align*}
But since for any $A,B$ $\det(1+AB)=\det(1+BA)$, we obtain
\begin{align*}
\det\mathcal{F}=&\det(I-\mathcal{S}\mathcal{D}\mathcal{S})
=\det\left(
\begin{array}{ll}I-\bar D^{(M)}\widehat{\mathcal{A}}^{(M)} &\bar D^{(M)}\mathcal{A}^{(M)}\\-\bar D^{(M)}\widehat{\mathcal{A}}^{(M)}
 &I+\bar D^{(M)}\mathcal{A}^{(M)}
\end{array}\right)=:\det\mathcal{F}_1,\quad\end{align*}
\begin{align*}
\det\mathcal{F}_1=\det(I+\bar D^{(M)}\mathcal{A}^{(M)})\det\Big(1-\bar D^{(M)}\widehat{\mathcal{A}}^{(M)}
+\bar D^{(M)}\widehat{\mathcal{A}}^{(M)}(I+\bar D^{(M)}\mathcal{A}^{(M)})^{-1}
\bar D^{(M)}{\mathcal{A}}^{(M)}\Big)\\
=\det(1-\bar D(\widehat{\mathcal{A}}^{(M)}-\mathcal{A}^{(M)}))=\det(1-\bar D\widetilde{\mathcal{L}}^{(M)}).
\notag\end{align*}
Similarly, since $\mathcal{S}\mathcal{F}^{-1}\mathcal{S}=S^2\mathcal{F}^{-1}_1$ and
\[\left(
\begin{array}{ll}I-\bar D\widehat{\mathcal{A}}^{(M)} &\bar D^{(M)}\mathcal{A}^{(M)}\\ -\bar D^{(M)}\widehat{\mathcal{A}}^{(M)}
&I+\bar D^{(M)}\mathcal{A}^{(M)}
\end{array}\right)^{-1}=\mathcal{G}^{(M)}
\left( \begin{array}{ll}I+\bar D^{(M)}\mathcal{A}^{(M)}&-\bar D^{(M)}\mathcal{A}^{(M)}\\ \bar D^{(M)}\widehat{\mathcal{A}}^{(M)}
 &I-\bar D^{(M)}\widehat{\mathcal{A}}^{(M)}
\end{array}\right),\]
where $\mathcal{G}^{(M)}:=\Big(1-\bar D\widetilde{\mathcal{L}}^{(M)}\Big)^{-1}$, we have
\begin{align*}
\mathrm{Tr}(\mathcal{S}\mathcal{F}^{-1}\mathcal{S}E)=\mathrm{Tr}(\mathcal{S}^2\mathcal{F}^{-1}_1E)=
\mathrm{Tr}(\mathcal{S}^2E)\mathcal{G}^{(M)}=\widetilde{\mathcal{L}}^{(M)}\mathcal{G}^{(M)}.
\notag\end{align*}

Hence we obtain for $I_{\bar n}^{*}$ of (\ref{I_*})
\begin{align*}
I_{\bar n}^{*}=&\mathrm{det}^{1/2}\mathcal{G}^{(M)}\exp\Big\{
\frac{\beta}{8}\Big(\widetilde{\mathcal{L}}^{(M)}\mathcal{G}^{(M)}\bar r^{(M)},\bar r^{(M)}\Big)\Big\}
(1+O(n^{-\kappa})).\end{align*}
Using Proposition \ref{p:appr} and Lemma \ref{l:DL}, we
can now replace $\widetilde{ \mathcal{L}}^{(M)}$ by the "block" integral
operator $\widetilde L$ with zero diagonal blocks
and off-diagonal blocks $\mathcal{L}_{\alpha,\alpha'}:
L_2[\sigma_{\alpha',{2\varepsilon}}]\to
L_2[\sigma_{\alpha,{2\varepsilon}}]$
\[\widetilde{\mathcal{L}}_{\alpha,\alpha'}[f]=\big(\mathbf{1}_{\sigma_{\alpha,{2\varepsilon}}}\widetilde{\mathcal{L}}
\mathbf{1}_{\sigma_{\alpha',{2\varepsilon}}}\big)[f]\]
The error of this replacement is $O(e^{-c\log^2n})$. Hence,
\begin{align}\label{I_*.2}
I_{\bar n}^{*}=
&\mathrm{det}^{1/2}\mathcal{G}\exp\Big\{\dfrac{\beta}{8}\big(\mathcal{G}\bar Dh,h\big)-\frac{\beta}{8}\big(\bar Dh,h\big)
+\dfrac{\beta}{2}\Big(\mathcal{G}
\big( X^{-1/2}_{\bar n}+(\frac{2}{\beta}-1)\nu\big),\bar Dh\Big)
\\&+
\frac{\beta}{2}\Big(\widetilde{\mathcal{L}}\mathcal{G}
\big( X^{-1/2}_{\bar n}+(\frac{2}{\beta}-1)\nu\big),
\big( X^{-1/2}_{\bar n}+(\frac{2}{\beta}-1)\nu\big)\Big)\Big\}(1+O(n^{-\kappa})).
\notag\end{align}
 Moreover, since the operator $\bar D$ is defined on
$\sigma$ (see (\ref{D,L}) and (\ref{bar_D})) and $ X^{-1/2}_{\bar n}$
are also defined on $\sigma$,
one can see that the operator $\widetilde{\mathcal{ L}}$ appears
in (\ref{I_*.2}) in the combination $\mathbf{1}_\sigma\widetilde{\mathcal{ L}}\mathbf{1}_\sigma$, so starting from this moment
we assume that $\widetilde {\mathcal{L}}:\mathcal{H}\to\mathcal{H}$. Let us study
\[\psi_{\bar n}:=\mathcal{G} X^{-1/2}_{\bar n}\Rightarrow
 X^{-1/2}_{\bar n}=(1-\bar D\widetilde{\mathcal{L}})\psi_{\bar n}.\]
In view of (\ref{DL})
we get
\begin{align*}
\widehat{\mathcal{L}} X^{-1/2}_{\bar n}=&\widehat{\mathcal{L}}(1-\bar D\widetilde{\mathcal{L}})\psi_{\bar n}=
\widehat{\mathcal{L}}\psi_{\bar n}+\widetilde{\mathcal{L}}\psi_{\bar n}-
(\widetilde{\mathcal{L}}\psi_{\bar n},X^{-1/2})\mathbf{1}_{\sigma_\alpha}
=\mathcal{L}\psi_{\bar n}+\mathrm{const}.
\end{align*}
Thus we conclude that
\[(\mathcal{L}\psi_{\bar n})_\alpha(\lambda)=c_\alpha(\bar n)=\mathrm{const}\Rightarrow
\psi_{\bar n}=\sum c_\alpha(\bar n)\psi^{(\alpha)},\]
where $\psi^{(\alpha)}$ are defined in (\ref{cond_psi}). Moreover, by (\ref{Q})-(\ref{cond_psi}) we
have
\begin{align*}
&\sum_{\alpha'}\mathcal{Q}_{\alpha\alpha'}c_{\alpha'}(\bar n)=(\psi_{\bar n},\mathbf{1}_{\sigma_\alpha})=
(\mathcal{G} X^{-1/2}_{\bar n},\mathbf{1}_{\sigma_\alpha})\\&=
( X^{-1/2}_{\bar n},\mathbf{1}_{\sigma_\alpha})-
(\widetilde{\mathcal{L}}\mathcal{G} X^{-1/2}_{\bar n},\bar D\mathbf{1}_{\sigma_\alpha})
=( X^{-1/2}_{\bar n},\mathbf{1}_{\sigma_\alpha})=\Delta n_\alpha\\
&\Rightarrow\quad\psi_{\bar n}=\sum_{\alpha,\alpha'}\mathcal{Q}_{\alpha\alpha'}^{-1}\psi^{(\alpha')}\Delta n_\alpha.
\end{align*}

Now let us transform the last two terms $S_3$ and $S_4$ in the r.h.s. of (\ref{I_*.2}).
\begin{align}\label{S_3}
S_3=&\frac{\beta}{2}\big(X^{-1/2}_{\bar n}+\big(\frac{2}{\beta}-1)\nu,\mathcal{G}^*(\widetilde{\mathcal{L}}\bar D-1+1)h\big)=
\frac{\beta}{2}\big(X^{-1/2}_{\bar n}+\big(\frac{2}{\beta}-1)\nu,\mathcal{G}^*h-h)\\
=&-\frac{\beta}{2}(X^{-1/2}_{\bar n},h)+\frac{\beta}{2}(\psi_{\bar n},h)-\big(\frac{\beta}{2}-1)\big((\mathcal{G}-1)\nu,h\big),
\notag\\\
S_4=&\frac{\beta}{2}(\widetilde{\mathcal{L}}\mathcal{G} X^{-1/2}_{\bar n}, X^{-1/2}_{\bar n})+
\frac{\beta}{2}\Big(\frac{2}{\beta}-1\Big)^2
\big(\widetilde{\mathcal{L}}\mathcal{G}\nu,\nu\big)
+2\big(\frac{\beta}{2}-1)\big(\widehat{\mathcal{L}}\psi_{\bar n},\nu\big),\notag\end{align}
since $(\widehat{\mathcal{L}}\psi_{\bar n},\nu\big)=
-(\widetilde{\mathcal{L}}\psi_{\bar n},\nu\big)$ in view of $({\mathcal{L}}\psi_{\bar n})_\alpha=\mathrm{const}$ and
$(\nu_\alpha,\mathbf{1}_{\sigma_\alpha})=0$.
Since by (\ref{DL}) $\widehat{\mathcal{L}}D=\widehat{\mathcal{L}}\bar D$, we obtain
\begin{align*}2\big(\widehat{\mathcal{L}}\psi_{\bar n},\nu_\alpha\big)=&
\big(\mathbf{1}_{\sigma_{\alpha}}\log  X_\alpha^{1/2},\psi_{\bar n}\big)-
\log(d_\alpha/2)\big(\mathbf{1}_{\sigma_\alpha},\psi_{\bar n}\big)-
(\widehat{\mathcal{L}}\bar D\log P_\alpha,\psi_{\bar n}\big)\\
=&\big(\mathbf{1}_{\sigma_{\alpha}}\log  X_\alpha^{1/2},\psi_{\bar n}\big)-
\log(d_\alpha/2)\big(\mathbf{1}_{\sigma_\alpha},\psi_{\bar n}\big)+\big(\log P_\alpha,\psi_{\bar n}\big)
-\big(X_{\bar n}^{-1/2},\log P_\alpha\big)\\
=&\big(\mathbf{1}_{\sigma_{\alpha}}\log \frac{\rho_\alpha}{\mu_\alpha},\psi_{\bar n}\big)-
\big(\mathbf{1}_{\sigma_{\alpha}}\log \frac{\rho_\alpha}{\mu_\alpha},X_{\bar n}\big).
\end{align*}
Thus
\begin{align}\label{S_4}
S_4=&\frac{\beta}{2}(\widetilde{\mathcal{L}}\mathcal{G} X^{-1/2}_{\bar n}, X^{-1/2}_{\bar n})+
\frac{\beta}{2}\Big(\frac{2}{\beta}-1\Big)^2
\big(\widetilde{\mathcal{L}}\mathcal{G}\nu,\nu\big)+\big(\frac{\beta}{2}-1)
\big(T,\psi_{\bar n}-X^{-1/2}_{\bar n}\big).
\end{align}
In addition, using that $\bar D\mathcal{L}\mathcal{G}X^{-1/2}_{\bar n}=0$, $\bar D\widehat{\mathcal{L}}X^{-1/2}_{\bar n}=0$,
we have
\begin{align*}
(\mathcal{L}\psi_{\bar n},\psi_{\bar n})=&(\mathcal{L}\mathcal{G}X^{-1/2}_{\bar n},\mathcal{G}X^{-1/2}_{\bar n})=
(\mathcal{L}\mathcal{G}X^{-1/2}_{\bar n},(1+\bar D\widetilde{\mathcal{L}}\mathcal{G})X^{-1/2}_{\bar n})\\=&
(\mathcal{L}\mathcal{G}X^{-1/2}_{\bar n},X^{-1/2}_{\bar n})=
((1+\bar D\widetilde{\mathcal{L}}\mathcal{G})X^{-1/2}_{\bar n},\widehat{\mathcal{L}}X^{-1/2}_{\bar n})+
(\widetilde{\mathcal{L}}\mathcal{G} X^{-1/2}_{\bar n}, X^{-1/2}_{\bar n})
\\=&(\widehat{\mathcal{L}} X^{-1/2}_{\bar n}, X^{-1/2}_{\bar n})+
(\widetilde{\mathcal{L}}\mathcal{G} X^{-1/2}_{\bar n}, X^{-1/2}_{\bar n}).
\end{align*}
This relation  combined with (\ref{I_*.2}), (\ref{S_3}), (\ref{S_4}), and (\ref{I_2}) yields
\begin{align*}\notag
I_{\bar n}^{(1)}I_{\bar n}^{*}=&\mathrm{det}^{1/2}\mathcal{G}\exp\Big\{\frac{2}{\beta}\big(\frac{\beta}{2}-1\big)^2
\big(\widetilde{\mathcal{L}}\mathcal{G}\nu,\nu\big)
+\dfrac{\beta}{8}\big(\mathcal{G}Dh,h\big)+
\Big(\frac{\beta}{2}-1\Big)\Big(\mathcal{G}\nu,h\Big)\Big\}\notag\\&
\cdot\exp\Big\{\dfrac{\beta}{2}\Big(\mathcal{L}\psi_{\bar n},\psi_{\bar n}\Big)
+\frac{\beta}{2}(\psi_{\bar n},h)
+\Big(\dfrac{\beta}{2}-1\Big)\big(\psi_{\bar n},T)\Big\}\big(1+O(n^{-\kappa})\big).
\end{align*}
Multiplying this by $I^{(0)}_{\bar n}$ from (\ref{I(u)}) and taking into account that
\begin{align*}\sum F_\beta(n_\alpha)-F_\beta(n)=&\big(\frac{\beta}{2}-1\big)\sum\big(n\mu_\alpha\log\mu_\alpha+
(n_\alpha-n\mu_\alpha)\log\mu_\alpha\big)\\&-c_\beta\sum\log\mu_\alpha-
c_\beta(q-1)\log n
+O\big(\frac{||\Delta n||^2}{n}\big),\end{align*}
  $\sum n_\alpha=n$, $\sum\mu_\alpha=1$, and  $n_\alpha$ under consideration satisfy
(\ref{b_n}), we obtain (\ref{as_Q}).

\section{Auxiliary results}\label{s:4}

\textit{Proof of Proposition \ref{p:appr}}. Assume that $k_\alpha\ge k_{\alpha'}$,
and prove that
\begin{equation}\label{p1.2}
|I_{k}(\lambda)|:=\bigg|  \mathbf{1}_{\sigma_{\alpha',2\varepsilon}}(\lambda)
 \int_{\sigma_{\alpha,2\varepsilon}}
  \log|\lambda-\mu|\frac{p^{(\alpha)}_{k}(\mu)}{|X^{1/2}_{\sigma_{\alpha,2\varepsilon}}(\mu)|}d\mu\bigg|\le Ce^{-2dk}.
\end{equation}
Then, using that
\[\int_{\sigma_{\alpha',2\varepsilon}}\frac{|p^{(\alpha)}_{k}(\lambda)|}
{|X^{1/2}_{\sigma_{\alpha,2\varepsilon}}(\lambda)|}d\lambda\le 1, \]
we obtain (\ref{p1.1}), since $k+k'\le2\max\{k,k'\}$.
Changing the variables in the integral in (\ref{p1.2}) $\mu=c_\alpha+
d_\alpha\cos x$ with $c_\alpha=\frac{1}{2}\big(a_\alpha+b_\alpha\big)$,
$d_\alpha=\frac{1}{2}\big(b_\alpha-a_\alpha+4\varepsilon\big)$, and integrating by parts, we obtain
\begin{align*}
I_{k}(\lambda)=&d_\alpha k^{-1}\int_{0}^\pi\frac{\sin x\sin k x}{z-\cos x}dx=
d_\alpha(2k)^{-1}\int_{0}^\pi\frac{\cos(k-1)x-\cos(k+1)x}{z-\cos x}dx\\=&
\frac{d_\alpha}{8k\pi i}\oint\frac{\zeta^{k-1}-\zeta^{k+1}+
\zeta^{-k+1}-\zeta^{-k-1}}{\zeta^2+1-2z\zeta}d\zeta\\=&
\frac{d_\alpha}{8k\pi i}\oint\frac{\zeta^{k-1}-\zeta^{k+1}}
{\zeta^2+1-2z\zeta}d\zeta=d_\alpha\frac{\zeta^{k-1}(z)-\zeta^{k+1}(z)}{4k\sqrt{z^2-1}},
\end{align*}
where $z=(\lambda-c_\alpha)d_\alpha^{-1}$,
$|z|>1+\delta_1$, $\zeta(z)=z-\sqrt{z^2-1}$, $|\zeta(z)|\le e^{-2d}$. This proves (\ref{p1.2}).$\square$

\medskip

\textit{Proof of Lemma \ref{l:pos}}.
Consider the Fourier transform $\widehat a(k)$ of $a(|\lambda|)$. Integrating by parts, it is easy to get that
\begin{align}\label{pos.1}
k\widehat a(k)=&\int_0^\infty a(\lambda)\sin k\lambda d\lambda
=\frac{16}{(kd)^3}-\frac{24\sin kd}{(kd)^4}+24\int_{kd}^\infty\frac{\sin
t}{t^5}dt\\
&=24\int_{kd}^\infty\frac{2t+t\cos t-3\sin t}{t^5} dt.
\notag\end{align}
Here the last equality can be obtained by the differentiation of the both parts with respect to $kd$.
Let us check that the numerator in the last integral is positive. Indeed, it is 0 at $t=0$, its derivative is
positive on $(0,\pi)$, and  it is evidently positive for $t\ge\pi$. Hence we get the first
assertion of the lemma. To prove the second assertion, let us note that, applying the Taylor  formula
 to the function $a_1(\lambda):=\lambda^{-1}-a'(\lambda)$ at the point $\lambda_0=d$, we get
$a_1(\lambda)=(\lambda-d)^4\xi^{-5}(\lambda)>0$,  and $a_1'(\lambda)<0$ for $0<\lambda<d$. Hence
the Fourier transform of $l(|\lambda|)-a(|\lambda|)$ is
\begin{align*}
\widehat l(|\lambda|)-\widehat a(|\lambda|)=&\frac{1}{k}\int_0^\infty a_1(\lambda)\sin k\lambda
d\lambda\\
=&\frac{1}{k^2}\sum_{j=0}^\infty\int_0^\pi \Big(a_1((t+2j\pi)/k )-a_1((t+(2j+1\pi ))/k\Big)\sin t dt>0.
\end{align*}

$\square$
\medskip

\textit{Proof of Lemma \ref{l:DL}}
It is easy to see that, to prove (\ref{DL.1}), it suffices to show that
\begin{equation}\label{DL.2a}
\widehat{\mathcal{S}}_{\alpha\alpha}\overline{D}^{(M)}_{\alpha}\widehat{\mathcal{S}}_{\alpha\alpha}\le (1-\delta_1)
\Leftrightarrow\widehat{\mathcal{A}}^{(M)}_{\alpha\alpha}\overline{D}^{(M)}_{\alpha}
\widehat{\mathcal{A}}^{(M)}_{\alpha\alpha}\le
(1-\delta_1)\widehat{\mathcal{A}}^{(M)}_{\alpha\alpha}.
\end{equation}
Fix   some $\alpha$ and denote
$A:=\widehat{\mathcal{A}}_{\alpha\alpha}$, $D:=\overline{D}_{\alpha}$
and $L:=\widehat{\mathcal{L}}_{\alpha}$ the complete matrices, corresponding to the above operators.
Write them as a block matrices
\[A=\left(\begin{array}{cc}A^{(11)}&A^{(12)}\\
A^{(21)}&A^{(22)}\end{array}\right),\,\,D=\left(\begin{array}{cc}D^{(11)}&D^{(12)}\\
D^{(21)}&D^{(22)}\end{array}\right), \,\,L=\left(\begin{array}{cc}L^{(11)}&L^{(12)}\\
L^{(21)}&L^{(22)}\end{array}\right),\]
such that
$A^{(11)}=:\widehat{\mathcal{A}}^{(M)}_{\alpha\alpha}$, $D^{(11)}=\overline{D}^{(M)}_{\alpha}$, and
$L^{(11)}=\widehat{\mathcal{L}}^{(M)}_{\alpha}$. Below we will use  the inequality valid for any
 block matrix $B\ge 0$
\begin{equation}\label{DL.2b}B=\left(\begin{array}{cc}B^{(11)}&B^{(12)}\\
B^{(21)}&B^{(22)}\end{array}\right),\quad B^{(21)}(B^{(11)})^{-1}B^{(12)}\le B^{(22)}.
\end{equation}
Assume that we have proved the inequality
\begin{equation}\label{DL.2}
D\le (1-\delta_1)A^{-1}\quad\Leftrightarrow\quad
ADA\le (1-\delta_1)A\quad\Rightarrow \quad(ADA)^{(11)}\le (1-\delta_1)A^{(11)}.
\end{equation}

Then we get
\begin{align}\notag(A^{(11)}D^{(11)}A^{(11)}f,f)=&((ADA)^{(11)}f,f)-((ADA)^{(22)}f,f)-2\Re(A^{(12)}D^{(21)}A^{(11)}f,f)\\
\le&((ADA)^{(11)}f,f)-2\Re(A^{(12)}D^{(21)}A^{(11)}f,f).\label{DL.2d}\end{align}
But
\[|(A^{(12)}D^{(21)}A^{(11)}f,f)|\le ||(A^{(11)})^{-1/2}A^{(12)}D^{(21)}(A^{(11)})^{1/2}||\,(A^{(11)}f,f).\]
In addition, using (\ref{DL.2b}) for the matrix $A$, we get
\begin{align*}
&||(A^{(11)})^{-1/2}A^{(12)}D^{(21)}(A^{(11)})^{1/2}||^2\\=&
||(A^{(11)})^{1/2}D^{(12)}A^{(21)}(A^{(11)})^{-1}A^{(12)}D^{(21)}(A^{(11)})^{1/2}||\\\le&
||(A^{(11)})^{1/2}D^{(12)}A^{(22)}D^{(21)}(A^{(11)})^{1/2}||=
||(A^{(22)})^{1/2}D^{(21)}A^{(11)}D^{(12)}(A^{(22)})^{1/2}||.\end{align*}
Then,  taking into account that for
any small enough $\varepsilon>0$ (\ref{DL.2c}) implies that
$(-L^{(11)})\le(D^{(11)}+\varepsilon)^{-1}$, we can use (\ref{DL.2b}) for $D+\varepsilon$ in order to get
\[ D^{(12)}A^{(11)}D^{(12)}\le D^{(21)}(-L^{(11)})D^{(12)}\le
D^{(21)}(D^{(11)}+\varepsilon)^{-1}D^{(12)}\le D^{(22)}+\varepsilon.\]
Hence we obtain
\begin{align}\label{DL.4}
&||(A^{(11)})^{-1/2}A^{(12)}D^{(21)}(A^{(11)})^{1/2}||^2\le||(A^{(22)})^{1/2}D^{(22)}(A^{(22)})^{1/2}||\\\le&
\mathrm{Tr\,}(A^{(22)})^{1/2}D^{(22)}(A^{(22)})^{1/2}=\mathrm{Tr\,}A^{(22)}D^{(22)}.
\notag\end{align}
Integrating by parts it is easy to check that
\begin{align}\label{b_A}
k^2j^2A_{k,j}=&k^2j^2\int_0^\pi\int_0^\pi a\Big(d_q(\cos x-\cos y)\Big)\cos kx\cos jy \,dxdy\\=&
-d_q^3a'''(0)\int_0^\pi\sin^3x\cos kx\cos jx dx+\int_0^\pi\int_0^\pi
\tilde a(x,y)\cos kx\cos jy \,dxdy,
\notag\end{align}
where $d_q=\frac{1}{2}(b_q-a_q+4\varepsilon)$ and $\tilde a(x,y)$ is some bounded piece-wise continuous function.
Hence we conclude that
there exists a constant $C_0$ such that if we introduce the diagonal matrix $A_d$ with the entries
$(A_d)_{jk}=\delta_{jk}k^{-4}$, then
\[A_d^{-1/2}AA_d^{-1/2}\le C_0\quad\Rightarrow \quad A\le C_0A_d.\]
 Moreover, it is easy to check that there exists $C_1>0$ such that
\begin{align}\label{b_D} D^{(22)}_{kk}\le C_1k^2.\end{align}
Thus, from (\ref{DL.4}) and above bounds we obtain that
\[||(A^{(11)})^{-1/2}A^{(12)}D^{(21)}(A^{(11)})^{1/2}||^2\le C_0\mathrm{Tr\,}A^{(22)}_dD^{(22)}=
 C\sum_{k=M+1}^\infty k^{-2}\le O(M^{-1}).\]
Finally we have from (\ref{DL.2d}) and (\ref{DL.2})
 \begin{align*}(A^{(11)}D^{(11)}A^{(11)}f,f)&\le((ADA)^{(11)}f,f)+(A^{(11)}f,f)O(M^{-1/2})\\&\le
 (1-\delta_1+O(M^{-1/2}))(A^{(11)}f,f)\le(1-\delta_1/2)(A^{(11)}f,f). \end{align*}
Hence we need only to prove (\ref{DL.2}). Since the last relations of (\ref{DL}) yields
\begin{equation}\label{DL.2c}
(\overline{D}_\alpha v,v)=((-\widehat{\mathcal{L}}_{\alpha})^{-1}v,v)+\pi^{-2}(v,X_{\alpha}^{-1/2})^2
(\widehat{\mathcal{L}}_{\alpha}^{-1}\mathbf{1}_{\sigma_\alpha},\mathbf{1}_{\sigma_\alpha})
\le((-\widehat{\mathcal{L}}_{\alpha})^{-1}v,v),\end{equation}
 it suffices to prove that
\begin{equation}\label{DL.3}
(-\widehat{\mathcal{L}})^{-1}\le(1-\delta_1)\widehat{\mathcal{A}}^{-1}\Leftrightarrow
\widehat{\mathcal{A}}\le(1-\delta_1)(-\widehat{\mathcal{L}}).
\end{equation}
But the last bound is a corollary  of the following inequality for the Fourier transforms of $a(|\lambda|)$
and $\log|\lambda|^{-1}$
\[\widehat a(k)<(1-\delta_1)\widehat l(k)=(1-\delta_1)\pi/k.\]
Since we have already  proved this inequality for $\delta_1=0$ in Lemma \ref{l:pos}, we have $\widehat a(k)/\widehat l(k)<1$.
Besides,  it follows from (\ref{pos.1}) that
$\widehat a(k)\sim k^{-4}$, hence  $\widehat a(k)/\widehat l(k)\to 0$, as $k\to\infty$, and
moreover, $\widehat a(k)/\widehat l(k)\to 0$, as $k\to 0$. Thus there exists $\delta_1>0$ such that
\[\sup_{k>0}\widehat a(k)/\widehat l(k)=1-\delta_1.\]

To prove the first relation of (\ref{DL.1a}), we represent
\begin{align}\label{U}
U_1^c\subset &U_2\cup U_3\cup U_4\cup U_5,\\
U_2=&\{\bar u:\sum_\alpha(\widehat{\mathcal{S}}_\alpha D_{\alpha,\alpha}\widehat{\mathcal{S}}_\alpha u^{(1)}, u^{(1)})\le
 \frac{c_0}{2}\log n\wedge
 ({\mathcal{S}}_\alpha D_{\alpha,\alpha}{\mathcal{S}}_\alpha u^{(2)}, u^{(2)})\le \frac{c_0}{2}\log n\},\notag\\
U_3=&\{\bar u: \frac{c_0}{2}\log n\le\sum_\alpha
(\widehat{\mathcal{S}}_\alpha D_{\alpha,\alpha}\widehat{\mathcal{S}}_\alpha u^{(1)}, u^{(1)})\le n\log^2n\},\notag\\
U_4=&\{\bar u: (u^{(1)},u^{(1)})\le C_*n^2\wedge
\sum_\alpha(\widehat{\mathcal{S}}_\alpha D_{\alpha,\alpha}\widehat{\mathcal{S}}_\alpha u^{(1)}, u^{(1)})\ge n\log^2n\},\notag\\
U_5=&\{\bar u: (u^{(1)},u^{(1)})\ge C_*n^2\}.
\notag\end{align}
It is evident that
\begin{align*}Q_{ n_\alpha}[\mu_\alpha^{-1}V^{(a)}_\alpha-n_\alpha^{-1}\widetilde h_\alpha]\le&|\sigma_\alpha|^{n_\alpha}
\exp\Big\{\beta n^2_\alpha\max\{\mu_\alpha^{-1}V^{(a)}_\alpha-n_\alpha^{-1}\Re\widetilde h_\alpha|\}/2\Big\}\\\le&
|\sigma_\alpha|^{n_\alpha}\exp\Big\{\beta (n^2_\alpha C+n_\alpha\max\{|\Re\widetilde h_\alpha|\})/2\Big\}.
\end{align*}
Moreover,  the definition of $\widetilde h_\alpha$ (see (\ref{I(u)})) and the Schwarz inequality yield
\begin{align}\label{b_h}|\Re\widetilde h_\alpha|\le& C_1\Big(1+\max\Big|\sum_{j,k,\alpha'}
\widehat S_{j,\alpha';k,\alpha}u_{j,\alpha'}^{(1)}p_j^{(\alpha)}(\lambda)\Big|\Big)\\
\le &C_1\Big(1+|u^{(1)}|\max\sum_{j}|p_j^{(\alpha)}(\lambda)|\sum_{k,\alpha'}|\widehat S_{j,\alpha';k,\alpha}|^2\Big)\notag\\
\le&C_1+C_2|u^{(1)}|\sum_{j}\widehat{\mathcal{A}}_{jj}^{1/2}\le C_1+C_3|u^{(1)}|,
\notag\end{align}
where the last inequality is based on the fact that $\widehat{\mathcal{A}}_{jj}\le C j^{-4}$ in view of (\ref{b_A}).
 Hence, choosing $C_*=\beta C_3$, we obtain
\[\Big(\frac{\beta}{2\pi}\Big)^{Mq}\int_{U_5}e^{-\beta(\bar u,\bar u)/8}I(u)du\le e^{-n^2c}.\]
Similarly to (\ref{b_h}), we have
\begin{align*}|\Re\widetilde h_\alpha(\lambda_1)-\Re\widetilde h_\alpha(\lambda_2)|\le\sum_{j}|p_j^{(\alpha)}(\lambda_1)-
p_j^{(\alpha)}(\lambda_2)|\widehat{\mathcal{A}}_{jj}^{1/2}\\
\le C|u^{(1)}||\lambda_1-\lambda_2|^{1/2}\sum_{j}j^{1/2}
\widehat{\mathcal{A}}_{jj}^{1/2}\le C'|u^{(1)}||\lambda_1-\lambda_2|^{1/2}.
\end{align*}
Thus  $n^{-1}_\alpha\widetilde h_\alpha(\lambda)$ is a Holder function for $\bar u\in U_4$, and we can use the result
of \cite{BPS:95}, according to which
\begin{align*}&Q_{ n_\alpha}[\mu_\alpha^{-1}V^{(a)}_\alpha-n_\alpha^{-1}\Re\widetilde h_\alpha]
\\&\le
\exp\Big\{
\frac{\beta n_\alpha^2}{2}\max_{m\in\mathcal{M}_1^+[\sigma_{\alpha,\varepsilon}]}\{L[m,m]-(m,\mu_\alpha^{-1}V^{(a)}_\alpha
-n_\alpha^{-1}\widetilde h_\alpha)\}+Cn\log n\Big\},\end{align*}
where $\mathcal{M}_1^+[\sigma_{\alpha,\varepsilon}]$ is a set of positive unit measures with supports belonging to
$\sigma_{\alpha,\varepsilon}$.
Since
\[-\mu_\alpha^{-1}V^{(a)}_\alpha(\lambda)\le -2\mu_\alpha^{-1}L[\rho_\alpha](\lambda),\quad \lambda
\in\sigma_{\alpha,\varepsilon},\]
we have
\begin{align}\notag
&\max_{m\in\mathcal{M}_1^+[\sigma_{\alpha,\varepsilon}]}\{L[m,m]-(m,\mu_\alpha^{-1}V^{(a)}_\alpha
-n_\alpha^{-1}\Re\widetilde h_\alpha)\}\\\le&\max_{m\in\mathcal{M}_1^+[\sigma_{\alpha,\varepsilon}]}
\{L[m,m]-(m,2\mu_\alpha^{-1}L[\rho_\alpha]
-n_\alpha^{-1}\Re\widetilde h_\alpha)\}\notag\\
\le&\max_{m\in\mathcal{M}_1[\sigma_{\alpha,\varepsilon}]}\{L[m,m]-(m,2\mu_\alpha^{-1}L[\rho_\alpha]
-n_\alpha^{-1}\Re\widetilde h_\alpha)\}.\label{b_E}\end{align}
Here $\mathcal{M}_1[\sigma_{\alpha,\varepsilon}]$ is a set of all signed unit measures with supports belonging to
$\sigma_{\alpha,\varepsilon}$. It is easy to see that, if we remove the condition of positivity of measures,
then the maximum point $\rho_1$ is uniquely defined by the conditions:
\[2L[\rho_1](\lambda)-2\mu_\alpha^{-1}L[\rho_\alpha](\lambda)
-n_\alpha^{-1}\Re\widetilde h_\alpha(\lambda)=\mathrm{const},\quad\lambda\in
 \sigma_{\alpha,\varepsilon},\quad\int_{\sigma_{\alpha,\varepsilon}}\rho_1=1.\]
Hence $\rho_1=\mu_\alpha^{-1}\rho_\alpha+
\frac{1}{2}D_{\sigma_{\alpha,\varepsilon}}\widetilde h_\alpha$ and the r.h.s. of (\ref{b_E}) takes the form
\begin{align*}
E_\alpha(\bar u):=-\mu_\alpha^{-2}L[\rho_\alpha,\rho_\alpha]+
\frac{n_\alpha^{-2}}{4}(D_{\sigma_{\alpha,\varepsilon}}\Re\widetilde h_\alpha,\Re\widetilde h_\alpha)+n_\alpha^{-1}
(\widetilde h_\alpha,\mu_\alpha^{-1}\rho_\alpha).
\end{align*}
 But by the definition of $\widetilde h_\alpha$ (see (\ref{I(u)}))
\[n_\alpha^{-1}(\widetilde h_\alpha,\mu_\alpha^{-1}\rho_\alpha)=O(n^{-1}_{\alpha})
+O(n/n_\alpha-\mu_\alpha^{-1})=O(n^{-1}\log n).\]
Hence
\[E_\alpha(\bar u)=-\mu_\alpha^{-2}L[\rho_\alpha,\rho_\alpha]+
\frac{n_\alpha^{-2}}{4}(\widehat{\mathcal{S}}_\alpha D_{\sigma_{\alpha,\varepsilon}}\widehat{\mathcal{S}}_\alpha u^{(1)}, u^{(1)})
+O(n^{-1}\log n).\]
Thus,
\[I(u)\le \exp\{\frac{\beta}{8}(\widehat{\mathcal{S}}_\alpha D_{\sigma_{\alpha,\varepsilon}}
\widehat{\mathcal{S}}_\alpha u^{(1)}, u^{(1)})
+O(n\log n)\},\]
and using   the Chebyshev inequality and (\ref{DL.2a}), we can conclude that for small enough  $n$-independent $\tau>0$
\begin{align}\label{U_4}
&\Big(\frac{\beta}{2\pi}\Big)^{Mq}\int_{U_4}e^{-\beta(u,u)/8}I(u) du\\\le&\Big(\frac{\beta}{2\pi}\Big)^{Mq}
\int e^{-\beta(u,u)/8}I(u)
e^{\tau(\sum_\alpha(\widehat{\mathcal{S}}_\alpha D_{\alpha,\alpha}\widehat{\mathcal{S}}_\alpha u^{(1)}, u^{(1)})-
n\log^2n)}\notag\\
\le&\Big(\frac{\beta}{2\pi}\Big)^{Mq}
\int e^{-\beta(u,u)/8}e^{\frac{\beta}{8}(\widehat{\mathcal{S}}_\alpha D_{\sigma_{\alpha,\varepsilon}}
\widehat{\mathcal{S}}_\alpha u^{(1)}, u^{(1)})+O(n\log n)}e^{\tau(\sum_\alpha(\widehat{\mathcal{S}}_\alpha D_{\alpha,\alpha}\widehat{\mathcal{S}}_\alpha u^{(1)}, u^{(1)})-
n\log^2n)}\notag\\
=&C(\tau)e^{-\tau n\log^2 n+cn\log n}.
\notag\end{align}
Note that (\ref{DL.2a}) implies that the last integral is convergent, and since
$\widehat{\mathcal{S}}_\alpha D_{\sigma_{\alpha,\varepsilon}}\widehat{\mathcal{S}}_\alpha$ is a trace class operator
(see (\ref{b_A})--(\ref{b_D})),
 the additional quadratic form in the exponent  changes the integral only by some uniformly bounded multiplier.
For $u\in U_3$ (\ref{t2.Z}) implies
\begin{align*}
\frac{Q_{ n_\alpha}[\mu_\alpha^{-1}V^{(a)}_\alpha-n_\alpha^{-1}\Re\widetilde h_\alpha]}
{Q_{ n_\alpha}[\mu_\alpha^{-1}V^{(a)}_\alpha]}\le\exp\{
(\widehat{\mathcal{S}}_\alpha D_{\alpha,\alpha}\widehat{\mathcal{S}}_\alpha u^{(1)}, u^{(1)})+
O(n^{-1}M(u^{(1)}, u^{(1)})^{3/2})+C\}.
\end{align*}
Hence, similarly to (\ref{U_4}), for small enough  $n$-independent $\tau>0$ we have
\begin{align*}
&\Big(\frac{\beta}{2\pi}\Big)^{Mq}\int_{U_3}e^{-\beta(u,u)/8}I(u) du\\\le&\Big(\frac{\beta}{2\pi}\Big)^{Mq}\int e^{-\beta(u,u)/8}I(u)
e^{\tau(\sum_\alpha(\widehat{\mathcal{S}}_\alpha D_{\alpha,\alpha}\widehat{\mathcal{S}}_\alpha u^{(1)}, u^{(1)})-
c_0\log n)}=C(\tau)e^{-c_0\tau \log n/2}.
\end{align*}
By the same way we also obtain for $U_2$
\begin{align*}
&\Big(\frac{\beta}{2\pi}\Big)^{Mq}\int_{U_2}e^{-\beta(u,u)/8}I(u) du\\\le&\Big(\frac{\beta}{2\pi}\Big)^{Mq}\int e^{-\beta(u,u)/8}I(u)
e^{\tau(\sum_\alpha({\mathcal{S}}_\alpha D_{\alpha,\alpha}{\mathcal{S}}_\alpha u^{(2)}, u^{(2)})-
c_0\log n)}=C(\tau)e^{-c_0\tau \log n/2}.
\end{align*}
This completes the proof of the first bound of (\ref{DL.1a}). To prove the second bound we just use
the Chebyshev inequality like above for $U_2$ and $U_3$.  Lemma \ref{l:DL} is proved.
$\square$

\smallskip

{\bf Acknowledgements.} The author is grateful to  Prof. G.Chistyakov, who proposed the proof
of Lemma \ref{l:3}. The author would like  also to thank MSRI and the organizers of the semester
 "Random Matrix Theory and its Applications", where this work was started.

\end{document}